\documentclass[final,1p,times]{elsarticle}

\usepackage{multirow}%
\usepackage{amsmath,amssymb,amsfonts}%
\usepackage{amsthm}%
\usepackage{hyperref}
\hypersetup{
  colorlinks,
  citecolor=MidnightBlue,
  linkcolor=MidnightBlue,
  urlcolor=MidnightBlue}
\usepackage{mathrsfs}%
\usepackage[title]{appendix}%
\usepackage{xcolor}%
\usepackage{textcomp}%
\usepackage{manyfoot}%
\usepackage{booktabs}%
\usepackage{algorithm}%
\usepackage{algorithmicx}%
\usepackage{algpseudocode}%
\usepackage{listings}%
\usepackage{subcaption}
\usepackage{bm}
\usepackage{caption}
\usepackage{algorithm}
\usepackage{diagbox}
\usepackage{algpseudocode}
\usepackage{natbib}
\bibpunct{(}{)}{,}{a}{,}{,}
\usepackage[version=4]{mhchem}

\newtheorem{hypothesis}{Hypothesis}[section]

\newcommand\subfig[4]{
\begin{subfigure}[b]{#1\textwidth}
        \centering
         \includegraphics[width=#2\textwidth]{#3.png}
         \caption{#4}
         \label{#3}
\end{subfigure}
}
\newcommand{\ER}{Erdős-Rényi}
\newcommand{\kavg}[1]{\langle #1 \rangle}

\newcommand{\rev}[1]{\textcolor{black}{#1}}

\begin{document}

\begin{frontmatter}



\title{\rev{Reducing Size Bias in Epidemic Network Modelling}}


\author[inst1]{Neha Bansal\corref{cor1}}
\ead{bansaln3@cardiff.ac.uk}
\cortext[cor1]{Corresponding author.}

\affiliation[inst1]{organization={School of Mathematics, Cardiff University},
            addressline={Senghennydd Road}, 
            city={Cardiff},
            postcode={CF24 4AG}, 
            country={UK}}

\author[inst1]{Katerina Kaouri}
\ead{KaouriK@cardiff.ac.uk}
\author[inst1]{Thomas E. Woolley}
\ead{WoolleyT1@cardiff.ac.uk}

\begin{abstract}
\rev{Epidemiological models help policymakers mitigate disease spread by predicting transmission metrics based on disease dynamics and contact networks. Calibrating these models requires representative network sampling. We investigate the Random Walk (RW) and Metropolis-Hastings Random Walk (MHRW) algorithms for three network types: Erdős–Rényi (ER), Small-world (SW), and Scale-free (SF). Disease transmission is simulated using a stochastic susceptible-infected-recovered (SIR) framework. For ER and SW networks, RW overestimates infected individuals and secondary infections by $25\%$ due to size bias, favouring highly connected nodes. MHRW, though more computationally intensive, reduces size bias and provides more representative samples. For time-to-infection, both algorithms provide representative estimates. However, neither algorithm samples SF networks representatively, exhibiting significant variability. Furthermore, removing duplicate sampled nodes reduces MHRW’s accuracy across three network types. We apply both algorithms to a cattle movement network of $46,512$ farms combining ER, SW, and SF features. RW overestimates infected farms by about $100\%$ and secondary infections by over $900\%$, reflecting significant size bias, while MHRW estimates align within $1\%$ of the cattle network values. RW underestimates time-to-infection by about $40\%$, while MHRW overestimates it by $10\%$. Accuracy, again, deteriorates when duplicate nodes are removed. Our findings guide algorithm selection and intervention strategies based on network structure and disease severity; RW’s conservative estimates suit high-mortality, fast-spreading epidemics, while MHRW enables more precise interventions for slower epidemics.}
\end{abstract}


\begin{highlights}
\item \rev{Compared Random Walk and Metropolis-Hastings Random Walk in estimating disease metrics.}
\item \rev{Sampled \ER~(ER), Small-world (SW), Scale-free (SF) networks.}
\item \rev{Explored reduction of size bias, the over-representation of highly connected nodes.}
\item \rev{MHRW outperformed RW for ER and SW networks.}
\item \rev{RW suits fast and/or severe epidemics needing urgent interventions.}
\item \rev{MHRW suits slower and low-severity epidemics.}
\end{highlights}


\end{frontmatter}


\section{Introduction}

During an epidemic, accurately estimating metrics such as the number of infected individuals (epidemic size), effective reproduction number (secondary infections), and time-to-infection is crucial for effective disease mitigation \citep{nunes2024redefining}. Epidemic modelling provides a suite of tools, frameworks, and methodologies to generate disease metric estimates \citep{vynnycky2010introduction}. However, estimating such metrics on networks is challenging due to the complexity of contact networks \citep{avraam2025impact}, the chosen data sampling method \citep{danon2011networks}, and other factors, such as data quality. In particular, inconsistency between the sampling method and the underlying contact network can lead to biased and non-representative samples, skewing disease metric estimates.

\rev{In this work, we investigate how to reduce a specific form of sampling bias known as size bias \citep{arratia2019size}, where individuals with a higher number of contacts are disproportionately overrepresented in the sample. This bias can lead to severely incorrect estimation of disease spread metrics and ineffective interventions. We aim to reduce size bias by tailoring sampling algorithms to the epidemic contact network. Our approach focuses on statistically comparing two traversal-based sampling (TBS) methods, Random Walk (RW) \citep{wei2004towards} and Metropolis-Hastings Random Walk (MHRW) \citep{hu2013survey}, to evaluate their influence on disease metric estimation. We opt to use a stochastic susceptible-infected-recovered (SIR) model for disease spread on three network types: \ER~(ER), Small-world (SW), and Scale-free (SF). Additionally, we demonstrate the real-world applicability of our methodology by analysing cattle movement data from the British Cattle Movement Service (BCMS) (Personal Communication, June, 14, 2024), where size bias may affect policymaking.}

TBS methods are widely used for sampling networks \citep{gjoka2010walking, craft2011network, malmros2016random, fournet2017estimating, cui2022survey}. These methods begin with one or more initial nodes (called seeds) and, based on specific information about their neighbours, choose the next node to sample. The primary distinction among TBS methods lies in how the next node is selected. The RW and MHRW are classic and frequently used TBS algorithms. 

The RW algorithm is often used in epidemiological studies of large populations \citep{milligan2004comparison,qi2023efficient} as it is relatively computationally cheap compared to other sampling algorithms. Specifically, for infectious diseases, the RW algorithm works well for homogeneous contact networks, where individuals have a similar number of contacts. However, real-world contact networks are often heterogeneous \citep{nielsen2020heterogeneity}, meaning that individuals vary in their number of contacts. This heterogeneity can result in size-biased samples when using the RW algorithm \citep{stein2014online,li2015random}.
 
\rev{MHRW addresses the issue of size bias by adjusting the selection probability of each node based on its number of connections, thereby reducing the likelihood of sampling highly connected individuals. This correction allows MHRW to generate samples that more accurately reflect the degree distribution of the underlying network (UN). Although the MHRW algorithm is approximately $1.5-2$ times more computationally expensive compared to the RW algorithm \citep{bishop2006pattern}, it yields more representative samples, leading to more reliable statistical estimates of dynamics on heterogeneous contact networks. We note that the RW and MHRW algorithms are based on Markov chains and share the ``memoryless" property, leading to duplicates in the sample. Thus, we also compare and evaluate the effect of duplicate sample removal on the accuracy of disease metric estimates for the RW and MHRW algorithms. }

The extent and intensity of disease spread are driven broadly by three factors: the virus transmissibility, the recovery rate of individuals, and the population contact network \citep{nelson2014infectious}. Contact networks act as a means for infectious disease transmission. In a contact network, individuals are represented as nodes, and an edge represents the interaction between two nodes. The number of edges (connections) attached to a node is its degree \citep{newman2018networks}. In this work, we use undirected contact networks \citep{newman2018networks}, i.e., disease transmission can occur in both directions along an edge. 

We consider three types of contact networks:1) ER networks \citep{erdios1959random}, which serve as a simple yet useful benchmark for understanding disease spread in random, unstructured networks, for example, sexual contact networks \citep{holme2013extinction}; 2) SW networks \citep{watts1998collective}, which closely mimic social mixing networks, with high clustering and long-range links; and 3) SF networks \citep{li2005towards}, characterised by the presence of hubs with high connectivity, reflecting the presence of super-spreaders in a real-world contact network \citep{lieberthal2021connectivity}. SF networks exhibit significant heterogeneity of connections as found in real-world contact networks, whereas SW and ER networks have relatively uniform connectivity. These three theoretical networks capture many properties of real-world networks \citep{aldous2003graphs} and the diversity of behaviours that can arise.

\rev{We employ a mechanistic, stochastic Susceptible-Infected-Recovered (SIR) model \citep{Brauer2008, kermack1927contribution} to simulate disease spread on networks. The individual-level SIR model is formulated using stochastic equations that describe the rate of change in the probability of individuals being in a given disease state. We chose this model because it captures heterogeneous mixing patterns within the population, as determined by the contact network structure \citep{Kiss_Miller_Simon_2017}.}

In modelling disease spread, only a sample of the data is required. There are two key reasons for using sampling: a) the large cost and time needed in collecting data for an entire population and b) the high computational cost of processing large datasets. The accuracy of predictions of both statistical and mechanistic models depends on the quality of the sample; if the sample is not representative of the population contact network, then it leads to biased estimates of disease metrics \citep{suhail2021incorporating,joyal2022well}

Aside from size bias, various biases can arise from using inappropriate sampling methods, including demographic bias \citep{tyrer2016sampling}, geographic bias \citep{banerjee2010statistics}, and temporal bias \citep{kleinbaum1981selection}. For diseases which are highly transmissible via close contacts, such as COVID-19, measles, chickenpox, etc., the number of contacts is a critical factor in the spread of the virus, so it is essential to obtain an unbiased sample that accurately reflects the population’s contact distribution. Failure to do so can lead to size-biased predictions of the disease spread metrics and, consequently, to ineffective mitigation strategies. Size bias is also observed in other fields, such as survival analysis of cancer patients, where size bias can result in the over-representation of individuals with longer survival times. If uncorrected, size bias skews results, leading to an overestimation of survival times \citep{shen2009analyzing}.

\rev{Previous studies have shown that the RW algorithm yields size-biased samples when applied to various heterogeneous networks.} In \citep{gjoka2010walking}, the RW algorithm produces a size-biased sample, over-representing high-degree nodes, whereas the MHRW algorithm generates a sample with a degree distribution closely matching that of Facebook's network. Similar observations are reported in \citep{leskovec2006sampling} for various network types, including citation networks from the e-print arXiv repository \citep{leskovec2006sampling}, autonomous system networks \citep{oregon_route} (Internet router networks), affiliation networks in the Astrophysics category on arXiv \citep{leskovec2006sampling}, and trust networks from Epinions.com \citep{richardson2003trust}. \rev{However, there remains a gap in the epidemiological literature: despite the widespread use of network sampling in disease modelling, the comparative performance of RW and MHRW in estimating epidemic metrics has not been systematically assessed. To our knowledge, our study is the first to do so.}

This paper is structured as follows. In Section \ref{methods}, we quantify size bias in the estimation of disease metrics for the RW and MHRW algorithms, and in Section \ref{results} we compare key disease metrics such as number of infected individuals, the number of secondary infections, and the time-to-infection, for the three chosen network types (ER, SW, SF). Finally, in Section \ref{cattle data}, to illustrate the practical applications of our study, we analyse cattle movement network data provided by the British Cattle Movement Service (BCMS) (Personal Communication, June 14, 2024), which has hybrid properties related to the ER, SW and SF networks. We provide a summary and directions for future research in Section \ref{conclusion}.

\section{Methods}\label{methods}
For completeness, we briefly define and describe several commonly used properties of networks. We also describe the RW and MHRW algorithms used for sampling from networks. Additionally, we outline the formulation and simulation methodology of a stochastic SIR model for infectious disease spread on networks.

\subsection{Networks}\label{networktheory}
We consider contact networks \citep{bansal2010dynamic,craft2015infectious}. A contact network is defined by $G= \{V, E\}$, where $V$ is a set of node labels representing the individuals who can carry and transmit the disease, and $E$ is a set of edges that defines whether there have been interactions between the individuals. Let $N$ be the number of nodes in $G$, also known as network size. Let $k_v$ denote the degree of a node $v\in V$, which is defined as the number of edges (connections) of node $v$. Based on the network structure, we can define and calculate (at least numerically) $p_k$, the probability that a node is of degree $k$, as well as the average degree of the network $\kavg{k} $ \citep{newman2018networks}. We will randomly generate our network structures on the basis of these degree distributions. Namely, a number of nodes will be fixed, and the connections will be chosen to satisfy a given degree distribution.

\subsubsection{Network structures}\label{theo_net}
Real-world networks, such as the cattle movement network from the BCMS, which we analyse in Section \ref{cattle data}, rarely align perfectly with theoretical network structures like ER, SW, or SF networks. Instead, they often have a mixture of features from these networks, such as clustering and heterogeneous connectivity. To better understand the implications of these structural characteristics on disease spread, we first describe these three theoretical networks individually \citep{newman2018networks}. This approach helps us to determine how specific network features influence disease metrics, offering a baseline for understanding the more complex structure of the cattle movement network and other real networks.
\begin{figure}
    \centering
    \begin{minipage}{\textwidth}
        \centering
        \subfig{0.3}{1}{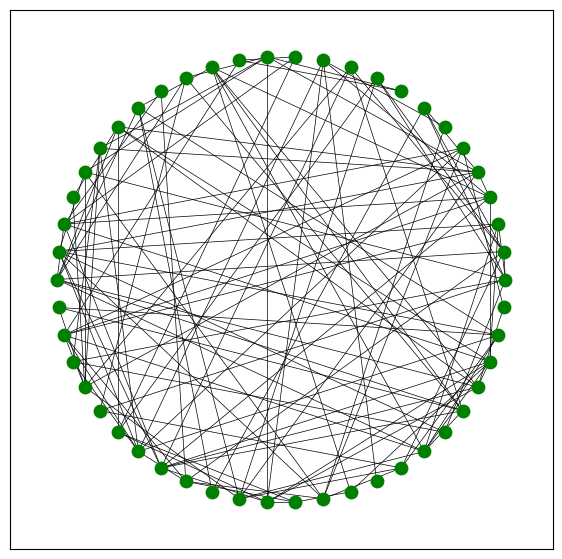}{\ER~network}
        \subfig{0.3}{1}{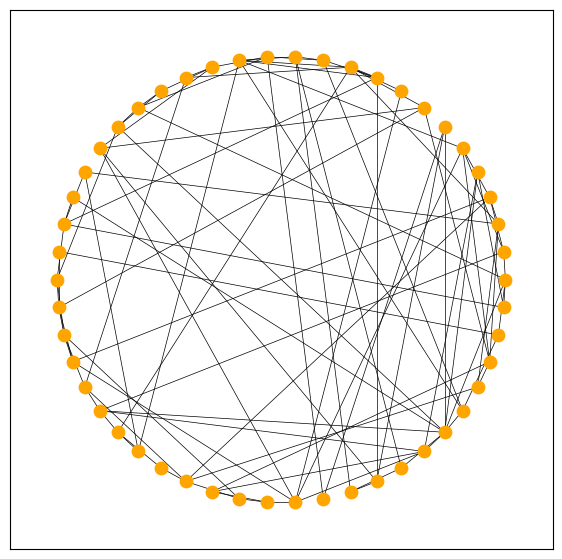}{Small-world network}
        \subfig{0.3}{1}{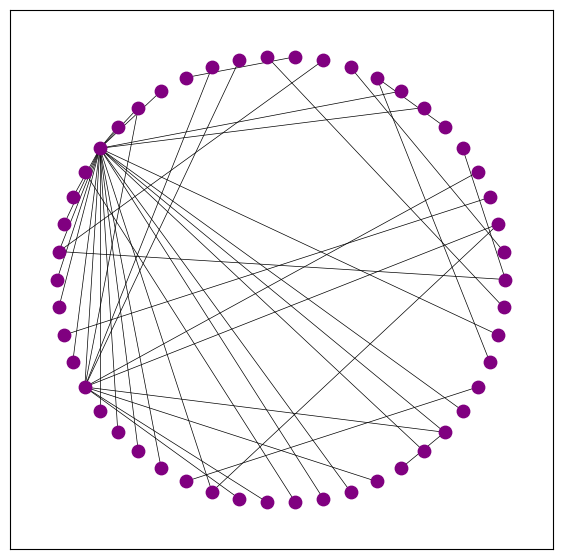}{Scale-free network}
        \caption{Illustration of three network structures, generated using the NetworkX library \citep{hagberg2008exploring}, with $N=50$ nodes and average degree $\langle k \rangle = 5$: a) ER network, b) SW network with $p=0.5$, and c) SF network with $\alpha=3$.}
        \label{Networks}
    \end{minipage}
    \begin{minipage}{\textwidth}
        \centering
        \subfig{0.3}{1}
        {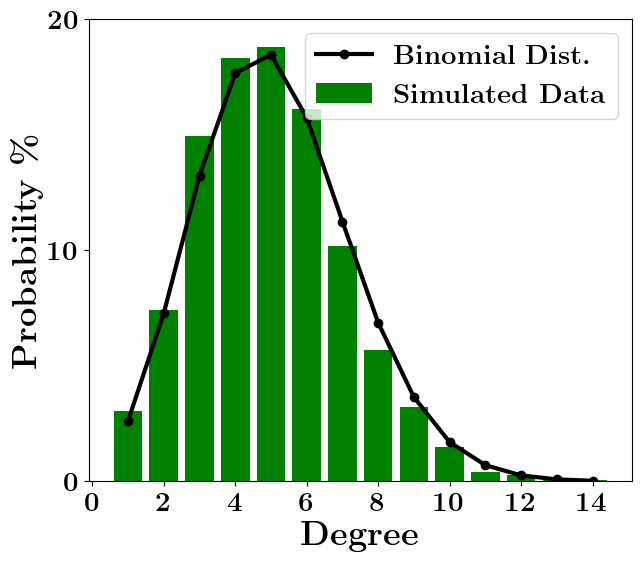}{\ER~network}
        \subfig{0.3}{1}
        {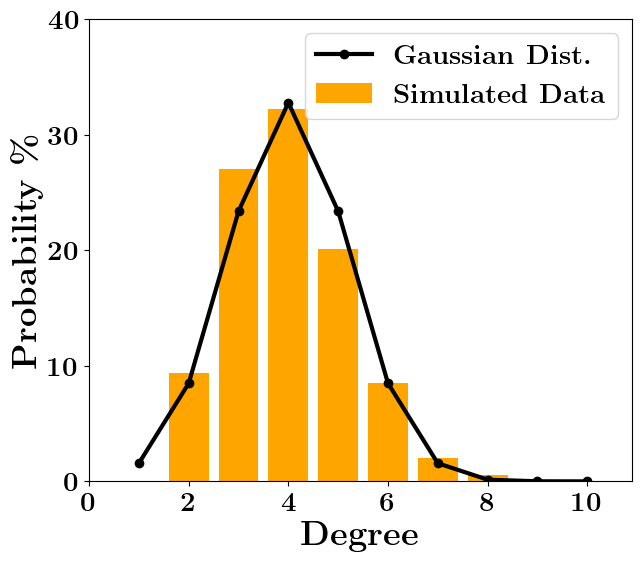}{Small-world network}
        \subfig{0.3}{1}
        {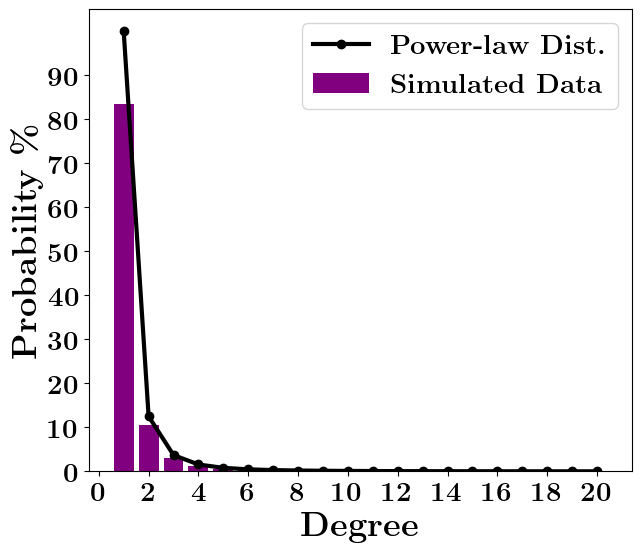}{Scale-free network}
        \caption{Degree distribution (histogram) and fitted theoretical degree distributions (black line plots): a) the binomial distribution given by Eq. \eqref{binom} with $N=50$ and $p=0.1$ for ER networks, b) a fitted Gaussian distribution with mean $4$ and standard deviation $1.2$, applied to the degree values from generated SW networks, and c) the power-law distribution given by Eq. \eqref{powerlaw} with $\alpha=3$ for SF networks. Degree distributions are estimated using $100$ simulated networks of each type with parameters $N=50$, $\langle k \rangle =5$ for the ER network, $p=0.5$ for the SW network, and $\alpha=3$ for the SF network, generated using the NetworkX library \citep{hagberg2008exploring}.}
        \label{Networks_deg}
    \end{minipage}
\end{figure}
\begin{enumerate}
    \item \textbf{\ER~(ER) networks} are generated by considering every pair of nodes and connecting them with an edge with probability $p$, independently of other edges in the network \citep{erdos1960evolution}.
    An example of an ER network is shown in Figure \ref{ER_net}. ER networks follow binomial degree distribution \citep{newman2001random}, with network size $N$ and edge creation probability $p=\langle k \rangle/(N-1)$ (Figure \ref{ER_net_deg}), that is, given by
    \begin{align}\label{binom}
        p_k = \binom{N}{k} p^k (1-p)^{N-k}.
    \end{align} 
    \item \textbf{Small-world (SW) networks} are characterized by high clustering and a small average shortest path distance between nodes (see Figure \ref{SW_net}), and they follow a skewed Gaussian degree distribution (Figure \ref{SW_net_deg}). Two common methods for generating SW networks are: 1) the Watts-Strogatz (WS) model \citep{watts1998collective}, and 2) the Newman-Watts-Strogatz (NWS) model \citep{Newman_Watts_1999}. In both methods, the process begins with a regular lattice network with average degree $\langle k \rangle$. In the NWS method, edges are added with probability $p$, whereas in the WS method, edges are rewired with probability $p$. For $p = 0$, the network is a regular lattice, and for $p = 1$, the network is an ER network. For $0 < p < 1$, a range of small-world networks with varying characteristics can be generated, and the value of $p$ can be selected based on the case. The example of an SW network shown in Figure \ref{SW_net} is generated using the WS method.
    \item \textbf{Scale-free (SF) networks} (see Figure \ref{SF_net}) follow a power-law degree distribution (Figure \ref{SF_net_deg}), indicating the presence of nodes with very high degree values, known as ``hubs,'' which lie in the tail of the distribution \citep{barabasi2003scale}. These hubs form the core of SF networks and play a critical role in maintaining network connectivity. The removal of a hub can cause significant disconnection in the network, as a disproportionate number of paths traverse through these hubs. An example of an SF network is shown in Figure \ref{SF_net}. 
    Typically, SF networks are generated using the preferential attachment process described in \citep{barabasi1999emergence}, wherein a new node preferentially connects to well-connected nodes. The probability of creating an edge to a node is proportional to its degree. The power-law degree distribution is expressed as:
    \begin{align}\label{powerlaw} p_k \propto k^{-\alpha}, \end{align}
    where $\alpha$ is the distribution parameter.
\end{enumerate}

\subsection{Sampling algorithms}\label{sample_algo} 
We compare the RW and the MHRW algorithms for sampling the ER, SW, and SF networks, as described in Section \ref{theo_net}. Below, we first describe the algorithms:
\begin{enumerate}
    \item \textbf{The RW algorithm} is a classic traversal-based sampling method \citep{gobel1974random}. Sampling starts with selecting an initial node at random from the network. The next node is chosen uniformly from the neighbours of the initial node. These two steps are repeated until the desired sample size is achieved. The probability of choosing a neighbour $w$ of node $v$ is denoted by $P_{v,w}^{RW}$, which is given by
    \begin{align}
        P_{(v,w)}^{RW} = \begin{cases}
            \frac{1}{k_v} &\text{ if } w \text{ is neighbour of } v,\\
            0 &\text{ otherwise},
        \end{cases}
    \end{align}
    where $k_v$ is the degree of node $v$.
    \item \textbf{The MHRW algorithm} is also a TBS method \citep{hu2013survey}. Sampling starts by randomly selecting a node from the network. The next node is selected from the neighbours of the initial node using the proposal-acceptance mechanism \citep{spencer2021accelerating} based on the degree of the two nodes. Suppose that the last sampled node is $v$ and the proposed next node is $w$, the probability of accepting $w$ is
    \begin{align}\label{MHRW_prob}
    P_{(v,w)}^{MH} = \begin{cases}
                    \frac{1}{k_v} \min\Big(1,\frac{k_v}{k_w}\Big) &\text{ if } w \text{ is neighbour of } v,\\
                    1- \sum_{y\neq v} P_{(v,y)}^{MH} &\text{ if  } w=v,\\
                    0 &\text{ otherwise}.
                    \end{cases}
    \end{align}
\end{enumerate}
Since the RW and MHRW are Markov chain-based algorithms and share the ``memoryless" property, some nodes may be revisited during the sampling process, leading to duplicates. The duplicates introduced by revisiting nodes reveal key structural features of the network. For example, they highlight community clustering, as random walks tend to revisit nodes within tightly connected groups \citep{rosvall2008maps}. Duplicates also emphasize high-degree nodes required for immunization and intervention strategies \citep{maiya2011benefits}. In biological and knowledge graphs, duplicates preserve important clusters and reflect the intensity of interactions, such as gene or drug-target relationships \citep{dempsey2012development,zhang2021discovering}.

\subsection{Stochastic SIR model}\label{sir formula}
In Section \ref{networktheory}, we introduced the structure and properties of the networks; we now demonstrate how to apply a stochastic SIR model on three networks (ER, SW, and SF) to compare disease metric estimates derived from samples obtained using the RW and MHRW algorithms.

Consider a population of size $N$, which is categorized into three compartments according to the infection status of the individuals during an epidemic. The three compartments are Susceptible ($S$), Infected ($I$), and Recovered ($R$). Let $\bm{X}(t) \equiv (S(t), I(t), R(t))$ be the state vector where $S(t), I(t), R(t)$ are the number of individuals in state $S, I, R$ respectively at time $t$. The number of individuals in each compartment of the SIR model changes due to two events: 1) a transmission event and 2) a recovery event, which can be written, respectively, as follows,
\begin{align}
\ce{$S$ + $I$ ->[$\beta/N$] $2I$} \label{SIR_r1},\\
\ce{$I$ ->[\gamma] $R$}\label{SIR_r2},
\end{align}
where $\beta/N$ is the transmission rate per infected individual per unit of time, and $\gamma$ is the recovery rate per infected individual per unit of time. 

The result of these two reactions/events is either the respective increase or decrease in the number of individuals in states $\{S, I, R\}$, which is translated as the change in the state vector $\bm{X}$. Let $\bm{v}_i \equiv (v_S,v_I,v_R)$ be the state-change vector, where $i=1$ denotes reaction \eqref{SIR_r1} and $i=2$ denotes reaction \eqref{SIR_r2}. The state change vector for reactions \eqref{SIR_r1} and \eqref{SIR_r2} are  $\bm{v}_1 = (-1,1,0)$, and $\bm{v}_2 = (0,-1,1)$, respectively.

The time-dependent state vector $\bm{X}(t)\equiv (S(t), I(t), R(t))$ is a Markov process, i.e., the state of the system at time $t+dt$, where $dt\geq 0$, is dependent only on the state at time $t$. Thus, the time evolution of $P(\bm{X},t | \bm{X}_0,t_0) \equiv$ probability that $\bm{X}(t) = \bm{X}$ given that $\bm{X}(t_0) = \bm{X}_0$ for any time $t\geq t_0$ is given below, which is the Chemical Master Equation (CME) \citep{gillespie1992rigorous,von2023computational}.
\begin{align}\label{CME}
   \frac{\partial P(\bm{X},t | \bm{X}_0,t_0)}{\partial t} &= \sum_{j=1}^2 \Bigg[ \underbrace{a_j(\bm{X}-\bm{v}_j) P(\bm{X}-\bm{v}_j, t | \bm{X}_0,t_0)}_{\text{inflow rate, probability to reach state } \bm{X} \text{ given that } \bm{X}(t) = \bm{X}-\bm{v}_j} \nonumber \\
   &\quad - \underbrace{a_j(\bm{X}) P(\bm{X},t|\bm{X}_0,t_0)}_{\text{outflow rate, probability to move away from state } \bm{X} \text{ given that } \bm{X}(t) = \bm{X}} \Bigg],
\end{align}
where $a_1(\bm{X}(t)) = S(t) I(t) \beta/N$, and $a_2(\bm{X}(t))=I(t) \gamma$ are the propensity functions.

Due to the nonlinearity of Eq. \eqref{CME}, obtaining a closed-form solution is not possible. However, we can simulate the SIR model on networks using the Gillespie algorithm \citep{Gillespie_2007}, which provides exact numerical realizations of the system dynamics.

\subsection{Network generation and stochastic SIR model simulations}\label{sim_s}
Above, we have discussed the theoretical aspects of network types, network properties, sampling algorithms, and a stochastic SIR model. In this section, we provide the libraries used for generating the networks and for simulating the SIR model on networks. The pseudocode for the RW and the MHRW sampling algorithms is in \ref{pseudocode}.

We use the NetworkX Python library \citep{hagberg2008exploring} to generate the three networks described in Section \ref{theo_net}. ER networks are generated using the \verb|gnp_random_graph| function, SW networks with the \verb|watts_strogatz_graph| function, and SF networks with the \verb|configuration| \verb|_model| function, where the input is a sequence of degree values drawn from a power-law distribution (using the \verb|random.zipf| function from the NumPy library \citep{2020NumPy-Array}). We generate $n=10,000$ networks for each type using the parameters listed in Table \ref{net_params_sim} to achieve a precision of $10^{-2}$ in the mean estimates of network characteristics, based on the relationship between sample size (or the number of simulations) and the standard error of the mean \citep{bondy1976standard}, given by 
\begin{align}
    \bar{\sigma}_x = \frac{\sigma}{\sqrt{n}},
\end{align}
where $\bar{\sigma}_x$ is standard error of the mean, $\sigma$ is the population standard deviation, and $n$ is the sample size.
\begin{table}
    \centering
    \begin{tabular}{|c|c|c|c|}
    \hline
     \backslashbox{\textbf{Network parameter}}{\textbf{Network type}} & \textbf{ER network} & \textbf{SW network} & \textbf{SF network} \\
     \hline
     Number of nodes & $10,000$ & $10,000$ & $10,000$ \\
     \hline
     Average degree & $5$ & $5$ & $5$ \\
     \hline
     Edge creation Probability $(p)$ & - & $0.5$ & - \\
     \hline
     Power-law exponent $(\alpha)$ & -  & - & $3$\\
     \hline
    \end{tabular}
    \caption{Chosen parameters for generating examples of three (ER, SW, SF) networks using the NetworkX library \citep{hagberg2008exploring}.}
    \label{net_params_sim}
\end{table}

In Figure \ref{Networks},  examples of an ER network, a SW network, and a SF network are shown, each generated using the NetworkX library with $50$ nodes, an average degree of $5$, and other parameters as mentioned in Table \ref{net_params_sim}.

We use the EoN Python library \citep{miller2020eon} to implement the Gillespie algorithm \citep{Gillespie_2007} for simulating a stochastic SIR model (Section \ref{sir formula}) on a network. The parameters for the SIR model include a recovery rate, $\gamma \in \{1, 1/7, 1/14\}$, and a transmission rate, $\beta/N$, which ranges from $0$ to $1$ with a step size of $0.1$. We first consider $\gamma = 1$ so that $\beta/N$ can be interpreted as a basic reproduction number for the SIR model. To understand the impact of longer recovery periods on estimates from samples, we use $\gamma = 1/7$ and $ 1/14$, which correspond to average recovery periods for influenza (one week) \citep{chowell2008seasonal} and COVID-19 (two weeks) \citep{seyedalinaghi2021predictors}, respectively.

After simulating the SIR model on the three networks, we obtain $100$ samples with $500$ nodes for each network type (ER, SW, and SF), using the RW and MHRW sampling algorithms (see pseudocode in Appendices \ref{alg:RW},\ref{alg:MHRW}). We generate $2500$ nodes per sample for each algorithm and discard the initial $2000$ nodes as burn-in nodes to allow the sampling algorithms to reach their stationary distribution and minimize the influence of the initial node on the resulting samples. Here, the sample size of $500$ (equivalent to $5\%$ of the population) is chosen to keep the margin of error approximately $\pm 5\%$ in the estimates \citep{conroy2016rcsi}. We obtain $100$ samples from each of the $10,000$ networks in each network type to get the precision of mean estimates up to the third decimal place \citep{bondy1976standard}. The code for the entire process pipeline is available on \href{https://github.com/nehabansal26/Epidemics-on-Networks}{GitHub}.

\subsection{Hypothesis test statements}\label{hyp_all}
We perform hypothesis testing to statistically compare the estimates of disease metrics in RW and MHRW samples relative to the underlying network (UN). The first step is to check if the data is normally distributed. We use one-sample Kolmogorov-Smirnov (KS) test \citep{massey1951kolmogorov} to test the following hypotheses \rev{for the ER, SW, and SF networks, respectively:}
\begin{hypothesis}\label{normal_test_pi}
\leavevmode\\
\textbf{Null Hypothesis (H\textsubscript{0})}: \\
The proportion of infected nodes for all $\beta$ values follows a Normal distribution. \\
\textbf{Alternative Hypothesis (H\textsubscript{1})}: \\
The proportion of infected nodes for all $\beta$ values does not follow a Normal distribution.
\end{hypothesis}
\begin{hypothesis}\label{normal_test_si}
\leavevmode\\
\textbf{Null Hypothesis (H\textsubscript{0})}: \\
The number of secondary infections for all $\beta$ values follows a Normal distribution. \\
\textbf{Alternative Hypothesis (H\textsubscript{1})}: \\
The number of secondary infections for all $\beta$ values does not follow a Normal distribution.
\end{hypothesis}
\begin{hypothesis}\label{normal_test_time_inf}
\leavevmode\\
\rev{
\textbf{Null Hypothesis (H\textsubscript{0})}: \\
Time-to-infection, for all $\beta$ values, follows a Normal distribution. \\
\textbf{Alternative Hypothesis (H\textsubscript{1})}: \\
Time-to-infection, for all $\beta$ values, does not follow a Normal distribution.}
\end{hypothesis}

Based on the results of the normality test, we then use the one-tailed Mann-Whitney U (U) test \citep{mann1947test} to test the hypotheses for comparing two samples and understand the direction of estimates from the samples relative to the UN. \rev{We test the following hypotheses for the ER, SW, and SF networks, respectively:}

\begin{hypothesis}\label{RW_manu_pi}
\leavevmode\\
\textbf{Null Hypothesis (H\textsubscript{0})}:
\begin{itemize}
    \item There is no difference between the probability distribution of the proportion of infected nodes in RW and UN or MHRW and UN.
    \item The proportion of infected nodes in UN is greater than or equal to those in RW or MHRW.
\end{itemize}
\textbf{Alternative Hypothesis (H\textsubscript{1})}: \\
The proportion of infected nodes in the UN tends to be less than those in RW or MHRW.
\end{hypothesis}
\begin{hypothesis}\label{MHRW_manu_pi}
\leavevmode\\
\textbf{Null Hypothesis (H\textsubscript{0})}:
\begin{itemize}
    \item There is no difference between the distribution of the proportion of infected nodes in RW and UN or MHRW and UN
    \item The proportion of infected nodes in UN is less than or equal to those in RW or MHRW.
\end{itemize}
\textbf{Alternative Hypothesis (H\textsubscript{1})}: \\
The proportion of infected nodes in the UN tends to be greater than those in RW or MHRW.
\end{hypothesis}
\begin{hypothesis}\label{RW_manu_si}
\leavevmode\\
\textbf{Null Hypothesis (H\textsubscript{0})}:
\begin{itemize}
    \item There is no difference between the probability distribution of the number of secondary infections in RW and UN or MHRW and UN
    \item Number of secondary infections in UN are greater than or equal to those in RW or MHRW.
\end{itemize}
\textbf{Alternative Hypothesis (H\textsubscript{1})}: \\
The number of secondary infections in the UN tends to be less than those in RW or MHRW.
\end{hypothesis}
\begin{hypothesis}\label{MHRW_manu_si}
\leavevmode\\
\textbf{Null Hypothesis (H\textsubscript{0})}:
\begin{itemize}
    \item There is no difference between the distribution of the number of secondary infections in RW and UN or MHRW and UN
    \item Number of secondary infections in UN are less than or equal to those in RW or MHRW.
\end{itemize}
\textbf{Alternative Hypothesis (H\textsubscript{1})}: \\
The number of secondary infections in the UN tends to be greater than those in RW or MHRW.
\end{hypothesis}

\begin{hypothesis}\label{RW_manu_time_inf}
\leavevmode\\
\rev{
\textbf{Null Hypothesis (H\textsubscript{0})}:
\begin{itemize}
    \item There is no difference between the probability distribution of time-to-infection in RW and UN or MHRW and UN
    \item Time-to-infection for UN is greater than or equal to that in RW or MHRW.
\end{itemize}
\textbf{Alternative Hypothesis (H\textsubscript{1})}: \\
Time-to-infection in the UN tends to be less than that in RW or MHRW.}
\end{hypothesis}
\begin{hypothesis}\label{MHRW_manu_time_inf}
\leavevmode\\
\rev{
\textbf{Null Hypothesis (H\textsubscript{0})}:
\begin{itemize}
    \item There is no difference between the distribution of time-to-infection in RW and UN or MHRW and UN
    \item Time-to-infection in UN is less than or equal to those in RW or MHRW.
\end{itemize}
\textbf{Alternative Hypothesis (H\textsubscript{1})}: \\
Time-to-infection in the UN tends to be greater than that in RW or MHRW.}
\end{hypothesis}

\section{Results}\label{results}
In this section, we present a comparative analysis of the RW and MHRW sampling algorithms, focussing on their ability to capture the degree distribution of the UN and on their impact on the estimation of disease metrics for three network types (ER, SW, and SF) relative to the UN. 

Since RW and MHRW are memoryless algorithms, they yield duplicate nodes in a sample \citep{hu2013survey}. We compare the outcomes when duplicates are removed or retained. This analysis aims to highlight how this subtle aspect can significantly influence the accuracy of disease metric estimation across network types.
 
\subsection{Degree distribution}\label{result_deg_dist}
\begin{figure}
    \centering
    \subfig{0.3}{1}{deg_dist_ratio_ER}{\ER~network}
    \subfig{0.3}{1}{deg_dist_ratio_SW}{Small-world network}
    \subfig{0.3}{1}{deg_dist_ratio_SF}{Scale-free network}
    \caption{The ratio of degree distributions for samples generated by the RW algorithm (red diamonds) and the MHRW algorithm (blue triangles), relative to the UN, is presented for three network types (ER, SW, SF). The UN degree distribution is based on $10,000$ networks of each type (the parameters are found in Table \ref{net_params_sim}) generated with the NetworkX library \citep{hagberg2008exploring}. Sample degree distribution is derived from $100$ samples (size $500$) for $10,000$ networks using RW and MHRW.}
    \label{deg_dist}
\end{figure}

In Figure \ref{deg_dist}, we compare the degree distribution ratios of the RW and MHRW samples to the UN for the ER, SW, and SF networks. The degree distribution for the UN is based on $10,000$ networks of each type, generated using the NetworkX Python library with parameters listed in Table \ref{net_params_sim}. For the samples, it is estimated from $100$ samples, each with a sample size of $500$, for $10,000$ networks of three types using the RW and MHRW algorithms. The vertical axis shows the probability ratio of a node with degree $k$ in the samples to that in the UN. The solid black line represents a ratio of $1$, so closeness to this line indicates that the sample's degree distribution is similar to the UN. Results are consistent for samples with and without duplicate nodes (data not shown).

For the ER (Figure \ref{deg_dist_ratio_ER}) and SW (Figure \ref{deg_dist_ratio_SW}) networks, where the average degree $\geq 5$, the MHRW samples (blue triangles) align closely with the UN, while the RW samples (red diamonds) increasingly deviate as degree values increase, showing size bias. This highlights the MHRW's accuracy in representing the UN's degree distribution.

In Figure \ref{deg_dist_ratio_SF}, both sampling methods do not perform well for the SF network. There is significant variability, and both methods over-represent high-degree nodes. The RW samples consistently show a higher ratio than the MHRW samples, showing more size bias. The presence of ``hubs" in SF networks leads to a high variance in the degree distribution, leading to higher variability in the ratio compared to the ER and SW networks. 

These results suggest that for networks with properties similar to the ER and SW, the MHRW sampling algorithm may be a more suitable choice for unbiased degree distribution. At the same time, networks with SF properties require a more delegated approach, such as Sampling for Large-Scale Networks at Low Sampling Rates (SLSR) \citep{Jiao_2024}, where core and peripheral nodes are sampled separately for balanced sampling.

\subsection{Disease metrics}\label{disesase_metrics}
In the previous section, we compared the degree distributions of the samples generated using the RW and MHRW algorithms. We observed that the degree distribution of MHRW samples is closer to that of the UN than that of RW samples due to a bias towards high-degree nodes. Now, we compare the sampling algorithms in estimating three disease metrics: 1) proportion of infected nodes, 2) average number of secondary infections, and 3) time-to-infection. We simulate the SIR model on networks to estimate the disease metrics in the samples and the UN. Here, we present the results for $\gamma=1$; see \ref{gammavar} for $\gamma \in \{1/7,1/14\}$.

\subsubsection{Proportion of infected nodes}\label{result_pi}
\begin{figure}
    \centering
    \begin{minipage}{\textwidth}
        \centering
        \caption*{Panel 1: Samples with duplicate nodes.}
        \subfig{0.3}{1}{ER_node_idx_ratio_SIR_ES}{\ER~network }
        \subfig{0.3}{1}{SW_node_idx_ratio_SIR_ES}{Small-world network}
        \subfig{0.3}{1}{SF_node_idx_ratio_SIR_ES}{Scale-free network}
    \end{minipage}
    \begin{minipage}{\textwidth}
        \centering
        \caption*{Panel 2: Samples without duplicate nodes.}
        \subfig{0.3}{1}{ER_node_idx_ratio_SIR_ES_dup}{\ER~network}
        \subfig{0.3}{1}{SW_node_idx_ratio_SIR_ES_dup}{Small-world network}
        \subfig{0.3}{1}{SF_node_idx_ratio_SIR_ES_dup}{Scale-free network}
    \end{minipage}
    \caption{The ratio of the proportion of infected nodes during an epidemic between RW (dashed red) and MHRW (dotted blue) samples, relative to their respective UNs, is shown for retained (1) and removed duplicate nodes  (2). The light red and blue bands indicate the range of one standard deviation. The UN infection count is based on SIR model simulations ($\gamma=1$) for $10,000$ networks of each type (the parameters are found in Table \ref{net_params_sim}) generated with the NetworkX library \citep{hagberg2008exploring}. Sample infection counts are derived from $100$ samples (size $500$) for $10,000$ networks using RW and MHRW.}
    \label{epi_size_ratio}
\end{figure}
In Figure \ref{epi_size_ratio}, we compare the proportion of infected nodes in the RW and MHRW samples relative to the UN as the transmission rate $\beta$ increases, with a recovery rate $\gamma=1$, for three networks (ER, SW, and SF). Panel 1 (Figure \ref{ER_node_idx_ratio_SIR_ES}, \ref{SW_node_idx_ratio_SIR_ES}, \ref{SF_node_idx_ratio_SIR_ES}) shows the results for samples with duplicate nodes, while Panel 2 (Figure \ref{ER_node_idx_ratio_SIR_ES_dup}, \ref{SW_node_idx_ratio_SIR_ES_dup}, \ref{SF_node_idx_ratio_SIR_ES_dup} ) shows results for samples without duplicates. The closer the ratio is to the solid black line, the closer the sample estimates are to those of the UN.

For $\beta > 0.2$, RW samples consistently overestimate the proportion of infected nodes in ER and SW networks compared to MHRW samples, regardless of the inclusion of duplicate nodes (Figure \ref{epi_size_ratio}). In SF networks, while MHRW samples tend to overestimate the average number of infected nodes relative to RW samples, estimates from both algorithms exhibit substantial variability and fail to represent the underlying network structure accurately.

For ER networks, the MHRW samples overestimate the proportion of infected nodes when duplicate nodes are removed (Figure \ref{ER_node_idx_ratio_SIR_ES_dup}) but align well when duplicate nodes are retained (Figure \ref{ER_node_idx_ratio_SIR_ES}). RW estimates remain consistent regardless of duplicate removal. Similar behaviour is observed for SW networks in Figure \ref{SW_node_idx_ratio_SIR_ES} and Figure \ref{SW_node_idx_ratio_SIR_ES_dup}.

For the SF networks, both sampling algorithms fail to sample the UN well (Figure \ref{SF_node_idx_ratio_SIR_ES}, Figure \ref{SF_node_idx_ratio_SIR_ES_dup}). In SF networks, removing duplicates increases the deviation between RW and MHRW estimates for $\beta > 0.6$, with MHRW overestimating the proportion of infected nodes relative to RW. High variability in SF networks reflects challenges in capturing the heterogeneity and complex structure (e.g., hubs and peripheral nodes; Figure \ref{deg_dist_ratio_SF}).

We use hypothesis testing to statistically compare the estimates from the RW and MHRW sampling algorithms \rev{for the ER, SW, and SF networks, respectively}. First, we conduct a one-sample KS test to assess whether the proportion of infected nodes follows a Normal distribution, as stated in Hypothesis \ref{normal_test_pi}, with significance level $\alpha = 0.05$. The KS test statistic is summarized in \ref{stat_results} Table \ref{infnodes} for three data groups: UN, RW, and MHRW for ER, SW, and SF networks. The KS statistic is approximately $0.5$ across networks and data groups, with all corresponding p-values less than $0.05$. These results indicate significant deviation from normality, providing evidence to reject $H_0$ in Hypothesis \ref{normal_test_pi}.

Since the proportion of infected nodes does not follow a Normal distribution, we conduct a one-tailed U-test to compare the RW and the MHRW sample estimates with the UN. We test two hypotheses, one for statistically significant overestimation, Hypothesis \ref{RW_manu_pi}, and another for statistically significant underestimation, Hypothesis \ref{MHRW_manu_pi}, with significance level $\alpha = 0.05$, \rev{for the ER, SW, and SF networks, respectively}. The U-test statistic provides a measure of closeness between the sample estimates and the UN, with a relatively larger value indicating greater similarity to the UN. The statistics of the U-test are summarized in Table \ref{U_test_ES} for samples with duplicate nodes and samples without duplicate nodes. 

The p-values are less than $0.05$ for three network types for the RW samples (with and without duplicates) and the MHRW samples without duplicate nodes, for Hypothesis \ref{RW_manu_pi}. This provides evidence to reject $H_0$ in Hypothesis \ref{RW_manu_pi} and indicates the significant overestimation of the proportion of infected nodes. For the MHRW samples with duplicate nodes, the p-value is less than $0.05$ for the SF network, evidence to reject $H_0$ in Hypothesis \ref{RW_manu_pi}. This indicates the significant overestimation by the MHRW samples for the SF network. On the other hand, for the MHRW samples with duplicate nodes in the ER and SW networks, the p-values are less than $0.05$ for Hypothesis \ref{MHRW_manu_pi} (reject $H_0$), indicating significant underestimation of the proportion of infected nodes.

We find that both sampling algorithms do not provide estimates statistically similar to the UN. However, for the MHRW sample estimates (with or without duplicates), the value of U-test statistics is greater than the RW sample estimates (see Table \ref{U_test_ES}) for the ER and SW networks. For example, in the case of the ER networks with duplicate nodes, U-test statistics for the MHRW is $5.90 \times 10^{11}$ and for the RW is $5.07 \times 10^{11}$. This indicates that the MHRW sample estimates are closer to the UN compared to the RW.

For SF networks, the RW sample estimates are closer to the UN compared to the MHRW, as the U-test statistics are higher for the RW. For example, in the case of duplicate nodes, the U-test statistic is $6.1 \times 10^{10}$ for RW and $2.8 \times 10^{10}$ for the MHRW. 

\begin{table}[h]
    \centering
    \begin{tabular}{|c|c|c|c|c|c|}
        \hline
        \backslashbox{\textbf{Network}}{\textbf{Sample}} & \multicolumn{2}{c|}{\textbf{With duplicate nodes}} & \multicolumn{2}{c|}{\textbf{Without duplicate nodes}} \\
        \hline
        & \textbf{RW} & \textbf{MHRW} & \textbf{RW} & \textbf{MHRW} \\
        \hline
        ER & $5.07 \times 10^{11} $ & $5.90\times 10^{11}$  & $5.11\times 10^{11}$ & $5.38\times 10^{11}$ \\
        \hline
        SW & $5.8\times 10^{11}$ & $5.9\times 10^{11}$ & $5.82\times 10^{11}$ & $5.86\times 10^{11}$ \\
        \hline
        SF & $6.1\times 10^{10}$ & $2.8\times 10^{10}$ & $5.72\times 10^{10}$ & $7.83\times 10^9$  \\
        \hline
    \end{tabular}
    \caption{U-test statistics (Hypotheses \ref{RW_manu_pi} and \ref{MHRW_manu_pi}) for comparing the estimates of the proportion of infected nodes in the samples generated using the RW and the MHRW sampling algorithms with the UN, with and without duplicate nodes.}
    \label{U_test_ES}
\end{table}
Overall, while MHRW is generally more accurate for ER and SW networks, its performance degrades for SF networks, where high variability and structural heterogeneity pose challenges for both algorithms.

\subsubsection{Average secondary infections}\label{result_si}
\begin{figure}
    \centering
    \begin{minipage}{\textwidth}
        \centering
        \caption*{(1) Samples with duplicate nodes.}
        \subfig{0.3}{1}{ER_second_inf_SIR_ES}{\ER~network }
        \subfig{0.3}{1}{SW_second_inf_SIR_ES}{Small-world network}
        \subfig{0.3}{1}{SF_second_inf_SIR_ES}{Scale-free network}
    \end{minipage}
    \begin{minipage}{\textwidth}
        \centering
        \caption*{(2) Samples without duplicate nodes.}
        \subfig{0.3}{1}{ER_second_inf_SIR_ES_dup}{\ER~network }
        \subfig{0.3}{1}{SW_second_inf_SIR_ES_dup}{Small-world network}
        \subfig{0.3}{1}{SF_second_inf_SIR_ES_dup}{Scale-free network}
    \end{minipage}
    \caption{The ratio of the number of secondary infections during an epidemic between RW (dashed red) and MHRW (dotted blue) samples, relative to their respective UNs, is shown for (1) retained and (2) removed duplicate nodes. The light red and blue bands indicate the range of one standard deviation. The UN estimates are based on SIR model simulations ($\gamma=1$) for $10,000$ networks of each type (the parameters are found in Table \ref{net_params_sim}) generated with the NetworkX library \citep{hagberg2008exploring}. Sample estimates are derived from $100$ samples (size $500$) for $10,000$ networks using RW and MHRW.}
    \label{second_inf}
\end{figure}
In Figure \ref{second_inf}, we compare the number of secondary infections (effective reproduction number) in the RW and MHRW samples relative to the UN  as the transmission rate $\beta$ increases, with a recovery rate $\gamma=1$, for three networks (ER, SW, and SF). Panel 1 (Figure \ref{ER_second_inf_SIR_ES}, \ref{SW_second_inf_SIR_ES}, \ref{SF_second_inf_SIR_ES}) shows the results for samples with duplicate nodes, while Panel 2 (Figure \ref{ER_second_inf_SIR_ES_dup}, \ref{SW_second_inf_SIR_ES_dup}, \ref{SF_second_inf_SIR_ES_dup}) shows results for samples without duplicates.

For $\beta>0.3$, RW samples (with or without duplicates) overestimate the number of secondary infections in ER and SW networks compared to the MHRW samples as shown in Figures \ref{ER_second_inf_SIR_ES}, \ref{SW_second_inf_SIR_ES}, \ref{ER_second_inf_SIR_ES_dup}, and \ref{SW_second_inf_SIR_ES_dup}. 

For ER networks, the removal of duplicate nodes results in overestimation by MHRW samples (Figure \ref{ER_second_inf_SIR_ES_dup}), while it aligns well when duplicate nodes are retained (Figure \ref{ER_second_inf_SIR_ES}). However, no change is observed for RW estimates in both scenarios (with or without duplicate nodes). Similar behaviour is observed for SW networks in Figure \ref{SW_second_inf_SIR_ES} and Figure \ref{SW_second_inf_SIR_ES_dup}.

For the SF network, both sampling algorithms overestimate the number of secondary infections with high variability (Figures \ref{SF_second_inf_SIR_ES} and  \ref{SF_second_inf_SIR_ES_dup}). However, when duplicate nodes are retained, MHRW sample estimates are closer to the UN than the RW estimates, whereas this behaviour reverses when duplicates are removed. We consistently observe that the network properties and disease metrics are highly variable and more sensitive to changes in the system parameters for the SF networks, indicating the limitation of both sampling algorithms to represent the UN well. 

We conduct hypothesis testing to statistically compare the estimates from the sampling algorithms. First, we conduct the one-sample KS test to assess whether the number of secondary infections follows a Normal distribution, as stated in Hypothesis \ref{normal_test_si}, with significance level $\alpha = 0.05$, \rev{for the ER, SW, and SF networks, respectively}. The KS test statistic is summarized in \ref{stat_results} Table \ref{secondinfs} for three data groups that are UN, RW, and MHRW for ER, SW, and SF networks. The KS statistic ranges from $0.5$ to approximately $0.6$ across networks and data groups, with all corresponding p-values less than $0.05$. These results indicate significant deviation from normality, providing evidence to reject $H_0$ in Hypothesis \ref{normal_test_si}.

Since the number of secondary infections does not follow a Normal distribution, we conduct a one-tailed U-test to compare the RW and the MHRW sample estimates with the UN, \rev{for the ER, SW, and SF networks, respectively}. We test two hypotheses, one for statistically significant overestimation (Hypothesis \ref{RW_manu_si}) and another for statistically significant underestimation (Hypothesis \ref{MHRW_manu_si}), with significance level $\alpha = 0.05$. The statistics of the U-test are summarized in Table \ref{U_test_SI_combined} for samples with duplicate nodes and samples without duplicate nodes.

The p-values are less than $0.05$ \rev{individually for three} network types for the RW samples (with and without duplicates) and the MHRW samples without duplicate nodes, for Hypothesis \ref{RW_manu_si}. This provides evidence to reject $H_0$ in Hypothesis \ref{RW_manu_si} and indicates a significant overestimation of the number of secondary infections. For the MHRW samples with duplicate nodes, p-values are less than $0.05$, \rev{respectively} for the ER and SF networks, evidence to reject $H_0$ in Hypothesis \ref{MHRW_manu_si}, implying a significant underestimation of the number of secondary infections. However, for the SW network, the p-value is greater than $0.05$ (accept $H_0$) for the MHRW sample estimates with duplicate nodes, for Hypotheses \ref{MHRW_manu_si} and \ref{RW_manu_si}, indicating that the MHRW estimates are statistically similar to the UN.

Similar to the proportion of infected nodes, in the case of the number of secondary infections, the U-test statistics (see Table \ref{U_test_SI_combined}) are higher for the MHRW sample estimates (with or without duplicates) compared to the RW for the ER and SW networks, implying that MHRW estimates are closer to the UN. For the SF networks, the RW sample estimates are closer to the UN compared to the MHRW, indicated by the higher values of U-test statistics in Table \ref{U_test_SI_combined}.

\begin{table}
    \centering
    \begin{tabular}{|c|c|c|c|c|}
        \hline
        \backslashbox{\textbf{Network}}{\textbf{Sample}} & \multicolumn{2}{c|}{\textbf{With duplicate nodes}} & \multicolumn{2}{c|}{\textbf{Without duplicate nodes}} \\
        \hline
        & \textbf{RW} & \textbf{MHRW}  & \textbf{RW} & \textbf{MHRW}  \\
        \hline
        ER & $2.5\times 10^{11}$ & $5.6\times 10^{11}$  & $2.5\times 10^{11}$ & $3.33\times 10^{11}$  \\
        \hline
        SW & $3.99\times 10^{11}$ & $6.0\times 10^{11}$ &$ 3.91\times 10^{11}$ & $4.51\times 10^{11}$  \\
        \hline
        SF & $8.2\times 10^{10}$ & $6.6\times 10^{10}$  & $8.45\times 10^{10}$ & $5.99\times 10^{10}$ \\
        \hline
    \end{tabular}
    \caption{U-test statistics (Hypotheses \ref{RW_manu_si} and \ref{MHRW_manu_si}) for comparing the estimates of the number of secondary infections in the samples generated using the RW and the MHRW sampling algorithms with the UN, with and without duplicate nodes.}
    \label{U_test_SI_combined}
\end{table}

\subsection{\rev{Time-to-infection}}\label{time_1}
\begin{figure}
    \centering
    \begin{minipage}{\textwidth}
        \centering
        \caption*{Panel 1: Samples with duplicate nodes.}
        \subfig{0.3}{1}{ER_inf_time_ratio_SIR_ES}{\ER~network }
        \subfig{0.3}{1}{SW_inf_time_ratio_SIR_ES}{Small-world network}
        \subfig{0.3}{1}{SF_inf_time_ratio_SIR_ES}{Scale-free network}
    \end{minipage}
    \begin{minipage}{\textwidth}
        \centering
        \caption*{Panel 2: Samples without duplicate nodes.}
        \subfig{0.3}{1}{ER_inf_time_ratio_SIR_ES_dup}{\ER~network}
        \subfig{0.3}{1}{SW_inf_time_ratio_SIR_ES_dup}{Small-world network}
        \subfig{0.3}{1}{SF_inf_time_ratio_SIR_ES_dup}{Scale-free network}
    \end{minipage}
    \caption{\rev{The ratio of the time-to-infection during an epidemic between RW (dashed red) and MHRW (dotted blue) samples, relative to their respective UNs, is shown for (1) retained and (2) removed duplicate nodes. The light red and blue bands indicate the range of one standard deviation. The UN estimates are based on SIR model simulations ($\gamma=1$) for $10,000$ networks of each type (the parameters are found in Table \ref{net_params_sim}) generated with the NetworkX library \citep{hagberg2008exploring}. Sample estimates are derived from $100$ samples (size $500$) for $10,000$ networks using RW and MHRW.}}
    \label{time_inf}
\end{figure}
\begin{table}
    \centering
    \begin{tabular}{|c|c|c|c|c|}
        \hline
        \backslashbox{\textbf{Network}}{\textbf{Sample}} & \multicolumn{2}{c|}{\textbf{With duplicate nodes}} & \multicolumn{2}{c|}{\textbf{Without duplicate nodes}} \\
        \hline
        & \textbf{RW} & \textbf{MHRW}  & \textbf{RW} & \textbf{MHRW}  \\
        \hline
        ER & $6.2\times 10^{11}$ & $5.98\times 10^{11}$  & $6.2\times 10^{11}$ & $6.08\times 10^{11}$  \\
        \hline
        SW & $6.1\times 10^{11}$ & $6.0\times 10^{11}$ &$ 3.91\times 10^{11}$ & $4.51\times 10^{11}$  \\
        \hline
        SF & $1.26\times 10^{11}$ & $7.37\times 10^{10}$  & $8.45\times 10^{10}$ & $5.99\times 10^{10}$ \\
        \hline
    \end{tabular}
    \caption{\rev{U-test statistics (Hypotheses \ref{RW_manu_time_inf} and \ref{MHRW_manu_time_inf}) for comparing the estimates of the time-to-infection of the UN with the samples generated using the RW and the MHRW sampling algorithms, with and without duplicate nodes.}}
    \label{U_test_time_to_infection}
\end{table}

\rev{In Figure \ref{time_inf}, we compare estimates of the time-to-infection from the RW and MHRW samples with the UN for increasing $\beta$ values and $\gamma=1$ for the three networks (ER, SW, and SF). In Panel 1 (Figure \ref{ER_inf_time_ratio_SIR_ES}, \ref{SW_inf_time_ratio_SIR_ES}, \ref{SF_inf_time_ratio_SIR_ES}) results for samples with duplicate nodes and in Panel 2 (Figure \ref{ER_inf_time_ratio_SIR_ES_dup}, \ref{SW_inf_time_ratio_SIR_ES_dup}, \ref{SF_inf_time_ratio_SIR_ES_dup}) results for samples without duplicate nodes are shown.}

\rev{We observe that for ER networks and $\beta<0.2$, there is a large variation in the estimates of both the sampling algorithms in both panels. For $\beta \geq 0.2$, the MHRW sample estimates align well with the UN, while the RW samples slightly underestimate the time-to-infection if duplicate nodes are retained in the samples (Figure \ref{ER_inf_time_ratio_SIR_ES}). When duplicate nodes are removed, the MHRW sample estimates slightly underestimate the time-to-infection (Figure \ref{ER_inf_time_ratio_SIR_ES_dup}), although the behaviour of the RW sample estimates remains consistent.} 

\rev{For SW networks, a behaviour similar to that of the ER networks is observed (Figure \ref{SW_inf_time_ratio_SIR_ES}, \ref{ER_inf_time_ratio_SIR_ES_dup}). However, the large variation in the estimates for both sampling algorithms persists up to $\beta=0.5$ for the SW networks.}

\rev{Interestingly, for SF networks, as shown in Figures \ref{SF_inf_time_ratio_SIR_ES} and \ref{SF_inf_time_ratio_SIR_ES_dup}, the behaviour of both algorithms remains consistent across both panels, i.e., estimates from both the algorithms align well with the UN and have a thin variation band. Overall, both sampling algorithms perform well across network types, particularly for the Scale-free (SF) networks, which is not the case for the other two disease metrics (the proportion of infected nodes and the number of secondary infections).}

\rev{We conduct hypothesis testing to statistically compare the estimates from the sampling algorithms. First, we conduct the one-sample KS test to assess whether the time-to-infection follows a Normal distribution, as stated in Hypothesis \ref{normal_test_time_inf}, with significance level $\alpha = 0.05$, for the ER, SW, and SF networks individually. The KS test statistic is summarized in \ref{stat_results} Table \ref{time_to_inf} for three data groups that are UN, RW, and MHRW for ER, SW, and SF networks. The KS statistic ranges from $0.5$ to approximately $0.7$ across networks and data groups, with p-values less than $0.05$ individually for all three networks. These results indicate significant deviation from normality, providing evidence to reject $H_0$ in Hypothesis \ref{normal_test_time_inf}.}

\rev{Since time-to-infection does not follow a Normal distribution, we conduct a one-tailed U-test to compare the RW and the MHRW sample estimates with the UN individually for the ER, SW, and SF networks. We test two hypotheses, one for statistically significant overestimation (Hypothesis \ref{RW_manu_time_inf}) and another for statistically significant underestimation (Hypothesis \ref{MHRW_manu_time_inf}), with significance level $\alpha = 0.05$. The statistics of the U-test are summarized in Table \ref{U_test_time_to_infection} for samples with duplicate nodes and samples without duplicate nodes.}

\rev{The p-values are less than $0.05$ individually for three network types, for the RW samples (with and without duplicates), and for Hypothesis \ref{MHRW_manu_time_inf}. This provides evidence to reject $H_0$ in Hypothesis \ref{MHRW_manu_time_inf} and indicates a significant underestimation of the time-to-infection.}

\rev{For the MHRW samples, without duplicate nodes individually for all three networks and with duplicate nodes for the SW and SF networks, p-values are less than $0.05$, evidence to reject $H_0$ in Hypothesis \ref{MHRW_manu_time_inf}, implying a significant underestimation of the time-to-infection. However, for the ER network, the p-value is less than $0.05$ (accept $H_0$ in Hypothesis \ref{RW_manu_time_inf}) for the MHRW sample estimates with duplicate nodes, indicating that the MHRW significantly overestimates the time-to-infection.}

\rev{In the case of time-to-infection, the U-test statistics (see Table \ref{U_test_time_to_infection}) are similar for the RW sample estimates (with or without duplicate nodes) compared to the MHRW for the ER and SW networks, implying that both RW and MHRW estimates are closer to the UN. However, for SF networks, the U-test statistics are higher for RW sample estimates (with or without duplicate nodes) compared to the MHRW, implying that RW estimates are closer to the UN.}

\rev{We observed that in Figure \ref{time_inf}, estimates from both sampling algorithms (with or without duplicates) are close to the UN. However, on average, the RW samples underestimate and the MHRW samples overestimate the time-to-infection across three network types.}

Overall, the MHRW algorithm is more representative of the UN compared to the RW in estimating disease metrics for ER and SW networks. Thus, if the UN for real-world networks is similar to the ER and SW networks, the MHRW algorithm is recommended for sampling. 

The RW samples overestimated \rev{the proportion of infected nodes and the number of secondary infections and underestimated the time-to-infection} regardless of whether duplicate nodes were removed or retained. The size bias in degree distribution (Section \ref{result_deg_dist}) for the RW samples leads to overestimating the number of secondary infections for all networks, as the nodes with high degrees are more likely to become infected. 

The MHRW samples align well with the UN for the ER and the SW networks for all disease metrics only when duplicate nodes are retained in the sample. The removal of duplicate nodes leads to an overestimation of the disease metrics (Section \ref{result_pi}, \ref{result_si} and \ref{time_1}). One reason for overestimation after removing duplicates for MHRW samples is the reduction in the number of unique nodes in the MHRW samples, with a comparatively smaller reduction in the number of unique infected nodes. This may occur because the MHRW algorithm tends to get `stuck' at nodes with very low connectivity due to fewer potential candidate nodes for the acceptance-rejection mechanism. Another contributing factor may be the relatively higher presence of highly connected nodes—which are more likely to become infected—in the MHRW samples compared to the UN, as observed in Figure \ref{deg_dist}.

Notably, for the SF network, both the sampling algorithms fail to obtain representative samples of the UN, resulting in high variation in estimates of the degree distribution and disease metrics. The SF network structure is complex compared to the ER and the SW networks; there is more heterogeneity in degree values. Also, the structure consists of two types of components: 1) core nodes of very high degree and 2) peripheral nodes of low degree, which constitute the majority of the network. This diversity leads to dependency on the starting node for both sampling algorithms, leading to high variability in sample estimates. 

From an application perspective, the RW sampling algorithm provides conservative estimates of disease metrics, which is beneficial for devising interventions in cases of infectious diseases with high fatality rates. For such diseases, e.g. COVID-19 and Ebola, it is preferable to contain transmission as early as possible; hence, overestimating the number of infected individuals and the effective reproduction number (i.e., the number of secondary infections) aids the implementation of timely prevention measures. However, for diseases with low fatality rates, policymakers have more time to develop policies aligned with available resources. Here, the MHRW sampling algorithm is preferable, as its estimates are closer to the UN structure. In this context, precise estimates of disease metrics are integral for the efficient use of time and resources to contain the disease spread.

Furthermore, the performance of any sampling algorithm depends heavily on the complexity of the UN structure and disease parameters (such as transmission and recovery rates). For complex network structures, such as SF networks, sample estimates tend to be highly variable and sensitive to disease parameters, indicating the importance of providing uncertainty intervals with estimates to policymakers and considering alternative sampling algorithms \citep{Leskovec_Faloutsos_2006}, depending on the availability of computational resources.

We recommend that data scientists working with policymakers examine the sensitivity of sampling algorithms to network properties and disease spread parameters before selecting an algorithm. Real-world contact networks combine features of various theoretical networks; thus, it is essential to identify which network properties should remain consistent in the sample, depending on the mechanism of disease spread. Using a combination of sampling algorithms is also advised to leverage the strengths of each method.

\section{Cattle Movement Network}\label{cattle data}
\begin{figure}
    \centering
    \subfig{0.45}{1}{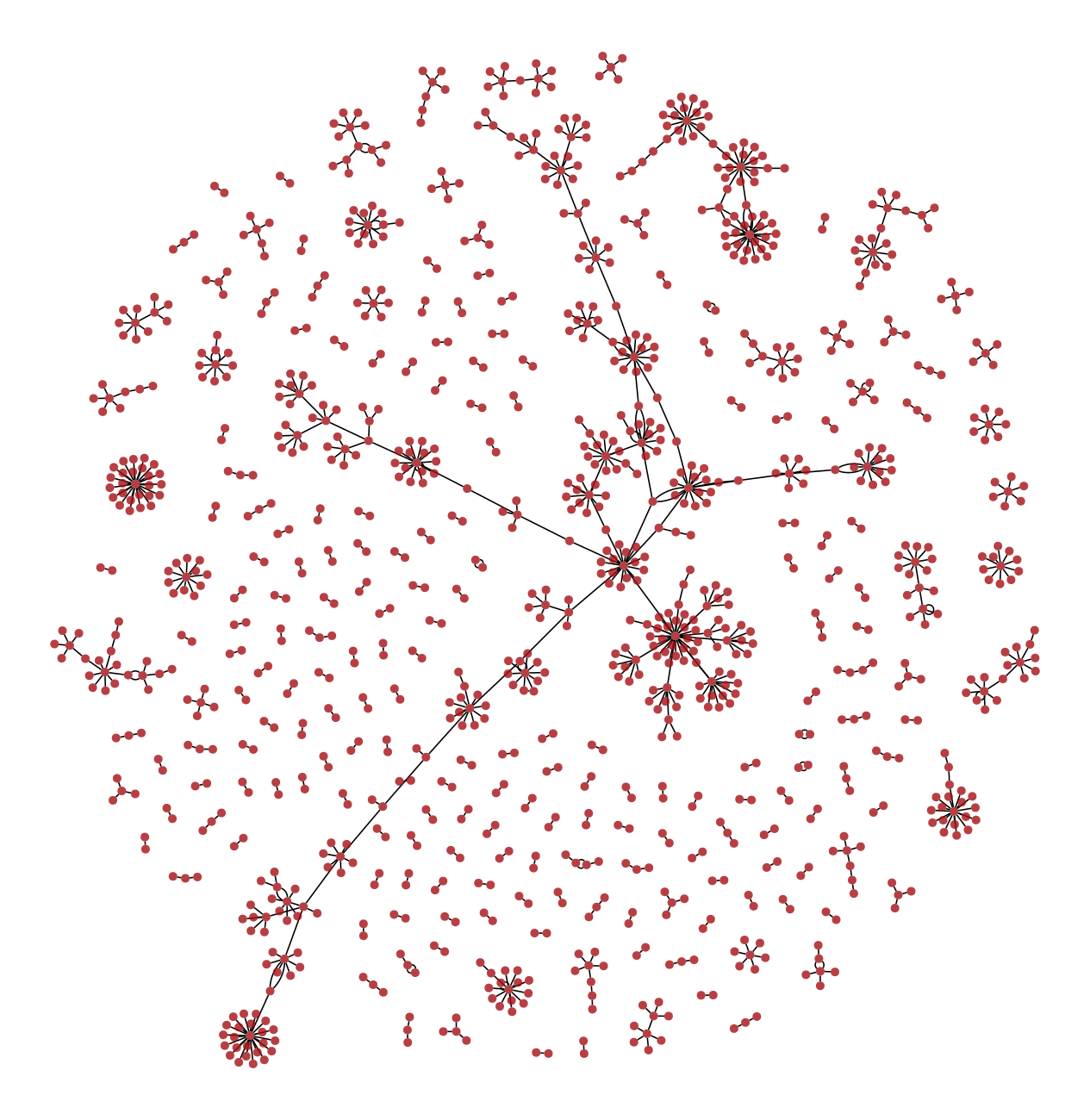}{Subset of the cattle movement network.}
    \subfig{0.45}{1}{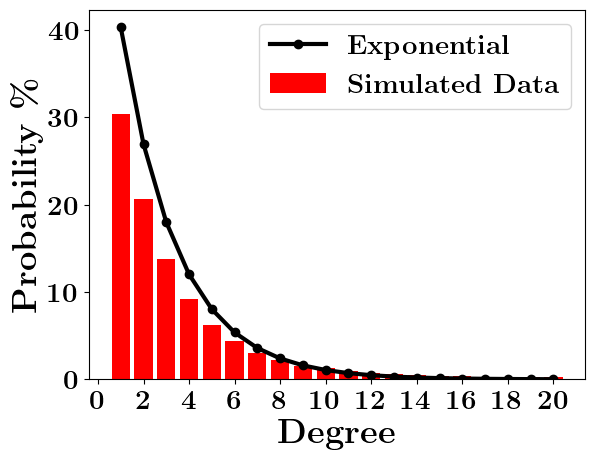}{Degree distribution.}
    \caption{Illustration of the cattle movement network data from BCMS.(a) A subset of the cattle movement network, where nodes represent individual farms and edges indicate recorded cattle movements between them.(b) Degree distribution of the cattle network. The red histogram represents the empirical data and the black curve shows an exponential fit for comparison.}
    \label{BCMS_deg_dist}
\end{figure}
In Section \ref{disesase_metrics}, we compared the network properties and disease metrics estimated from the RW and the MHRW sampling algorithms on three theoretical network types. Now, we implement both sampling algorithms on cattle movement network data. We utilise the cattle movement data provided by the British Cattle Movement Service (BCMS), which contains information on the birth, death, and movement of cattle. 

We use cattle movement data from January 2018 to March 2018, recorded at day level. We consider cattle holding places (farms, marketplaces, and slaughterhouses) as nodes and an edge is created between two nodes if there is cattle movement between them. Note that movement does not occur through all edges every day, which implies that the network structure changes over time. However, for this study, we assume that all edges are active and that the network is static.

\subsection{Data description}
The data contains movement details of $1,135,502$ cattle; there are two movement records on average per cattle. There are $46,512$ holding places, which are the nodes of the network. A static and undirected network is created with $46,512$ nodes and $159,036$ edges. Note that an edge between a pair of nodes is counted only once in case of multiple movement records. A part of the movement network is shown in Figure \ref{BCMS_graph_tools}; we observe a mix of large, connected components, small clusters of nodes, and isolated node pairs.

\begin{figure}
\centering
    \begin{minipage}{\textwidth}
    \centering
    \caption*{Panel 1: Samples with duplicate nodes.}
    \subfig{0.3}{1}{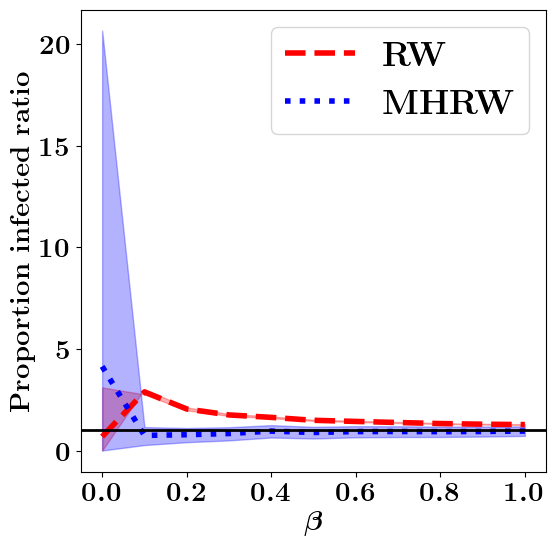}{Proportion of infected nodes}
    \subfig{0.3}{1}{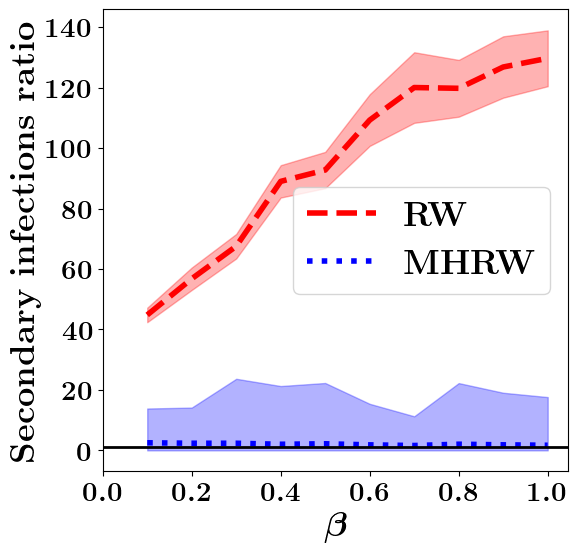}{Number of secondary infections}
    \subfig{0.3}{1}{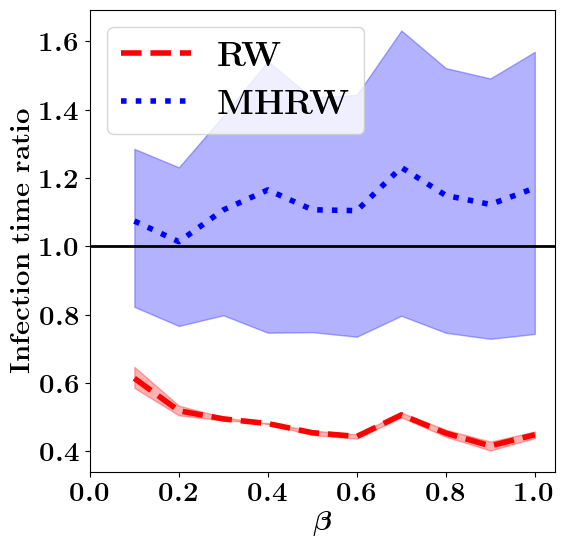}{Time-to-infection (days)}
    \end{minipage}

    \begin{minipage}{\textwidth}
    \centering
    \caption*{Panel 2: Samples without duplicate nodes.}
    \subfig{0.3}{1}{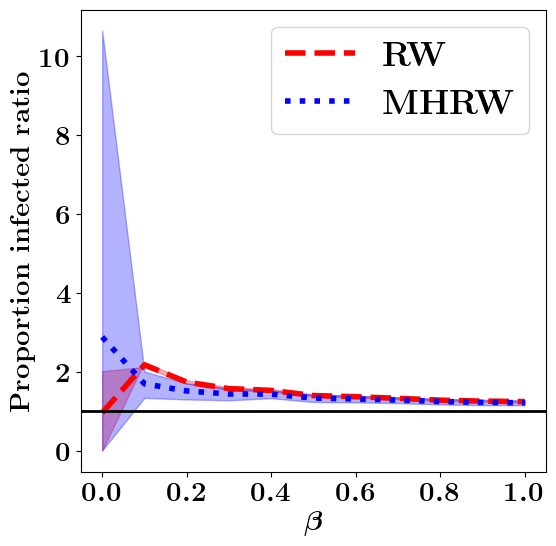}{Proportion of infected nodes}
    \subfig{0.3}{1}{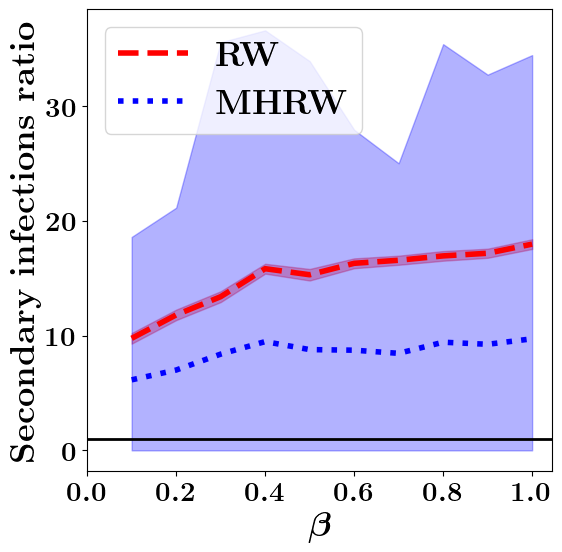}{Number of secondary infections}
    \subfig{0.3}{1}{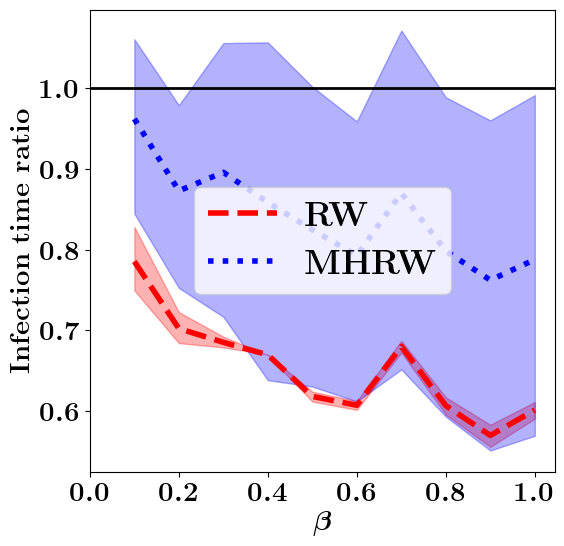}{Time-to-infection (days)}
    \end{minipage}
    \caption{The ratio of disease metric estimates between sampled generated using RW and MHRW algorithm, relative to the UN for two scenarios: (1) samples with duplicate nodes, and (2) samples without duplicate nodes. The light red and blue bands indicate the range of one standard deviation. disease metric estimates are based on $500$ SIR model simulations on the cattle movement network with $\gamma=1$. Sample estimates are derived from $10,000$ samples (size $2500$) for cattle movement network, using RW and MHRW.}
    \label{bcmssir}
\end{figure}
The network's degree ranges from $1$ to $6,556$, with an average degree of $6$. Notably, $30\%$ of the nodes have a degree of $1$, while only $2.69\%$ of the nodes have a degree greater than $20$. There is only one node with $6,556$ edges, which should be a marketplace.

The degree distribution for the network is shown in Figure \ref{deg_dist_BCMS}, up to a degree value of $20$. The degree distribution of the cattle network is fitted with an exponential distribution \citep{balakrishnan2019exponential} with rate parameter $0.4$ and location parameter $1$ (the distribution is shifted by one unit to the right along the horizontal axis), represented by the solid black line in Figure \ref{deg_dist_BCMS}. Other studies state that cattle movement network follows a power-law degree distribution \citep{christley2005network,fielding2019contact}. One reason for this difference is the amount and period of data; we have only three months of data (from 2018), whereas these studies analyse data for multiple years preceding 2018 \citep{duncan2022quantifying}.

We observed the presence of clusters with few connections outside the cluster, which is similar to the structure of the SW networks and agrees with the results in \citep{fielding2019contact}. We also observe a uniform connectivity pattern (average degree $2$), similar to the SW and ER networks, if we exclude nodes with one connection or a degree greater than $ 1000$. We observe the SF network's trait of the presence of ``hubs'' in the cattle movement network, which is suggested by the presence of $0.06\%$ nodes with degree $\geq 1000$ and $30\%$ nodes with degree $1$. There are a few nodes with very high degrees, likely markets, that facilitate extensive animal movement and can potentially act as key nodes for disease spread.

\subsection{Results}
Due to a lack of disease data, we study a hypothetical disease spread scenario, for example, bovine tuberculosis \citep{bekara2014modeling}. We run $500$ simulations of the stochastic SIR model on the cattle movement network using the methodology and parameters from Section \ref{sim_s}. The RW and MHRW sampling algorithms are used to generate $10,000$ samples, with a sample size of $2,500$ ($5\%$ of the network size). The number of samples, sample size, and number of SIR model simulations are chosen to be accurate to the second decimal place, as in Section \ref{results}.

In Figure \ref{bcmssir}, we compare three disease metrics in RW and MHRW samples relative to the cattle network, as the transmission rate $\beta$ increases, with a recovery rate $\gamma=1$, for the cattle movement network. The three disease metrics are: 1) proportion of infected nodes, 2) number of secondary infections, and 3) time-to-infection. Panel 1 (Figure \ref{bcms_network_node_idx_ratio_SIR_ES_right}, \ref{bcms_network_second_inf_SIR_ES_right}, \ref{bcms_network_inf_time_ratio_SIR_ES_right}) shows the results for samples with duplicate nodes, while Panel 2 (Figure \ref{bcms_network_node_idx_ratio_SIR_ES_dup}, \ref{bcms_network_second_inf_SIR_ES_dup}, \ref{bcms_network_inf_time_ratio_SIR_ES_dup}) shows those without duplicates. The closer the disease metric estimates are to the solid horizontal black line, the better the agreement to the underlying network metric values.

We observe that the estimate of the proportion of infected nodes in an epidemic is overestimated by RW samples in both panels (with and without duplicate nodes); see Figure \ref{bcms_network_node_idx_ratio_SIR_ES_right} and Figure \ref{bcms_network_node_idx_ratio_SIR_ES_dup}. However, the estimate from MHRW samples aligns very well with the cattle network when duplicate nodes are retained (Figure \ref{bcms_network_node_idx_ratio_SIR_ES_right}), but it overestimates when duplicate nodes are removed (Figure \ref{bcms_network_node_idx_ratio_SIR_ES_dup}). In the case of samples without duplicate nodes, the estimates of the RW and the MHRW samples are the same for $\beta>0.3$.

The RW samples overestimate the number of secondary infections when duplicate nodes are retained compared to the MHRW samples for the ER and SW networks, which align well with the cattle network (Figure \ref{bcms_network_second_inf_SIR_ES_right}). We observe a significant difference in the estimates when duplicate nodes are removed. The RW sample overestimates the number of secondary infections by a factor of 10–20, which increases to a factor of 40–130 when duplicate nodes are retained. In comparison, the MHRW sample without duplicates overestimates by a factor of approximately 7. Also, the variability in the estimates from the MHRW samples is higher when duplicate nodes are removed compared to RW samples. 

In Panel 1 of Figure \ref{bcmssir}, the estimate of time-to-infection is overestimated by the MHRW samples and underestimated by the RW samples. The variability in estimates is extremely high for the MHRW samples compared to the thin band for the RW sample estimates (Figure \ref{bcms_network_inf_time_ratio_SIR_ES_right}). Interestingly, after removing the duplicate nodes, both sampling algorithms underestimate the time-to-infection. This estimate deviates further below the cattle network estimate as the $\beta$ value increases.

Figure \ref{bcmssir} indicates that the estimates from MHRW samples are closer to the cattle network than the RW samples. However, the variability is significantly higher for the MHRW samples for the number of secondary infections and the time-to-infection compared to the RW samples.

The trend of disease metric estimates on the cattle movement network is consistent with those seen in the ER and SW networks. The RW samples provide conservative estimates of disease spread, whereas the MHRW samples yield estimates closer to the UN when duplicate samples are retained. For highly contagious diseases, such as Foot and Mouth Disease, which has a mortality rate of $20\%$ in young calves \citep{woah_fmd} and an average of $20$ secondary infections for direct transmission \citep{paton2018understanding}, the RW sampling algorithm is recommended. Conversely, for diseases such as Bovine Viral Diarrhoea, with a mortality rate of approximately $7\%$ \citep{dobos2024control} and the average number of secondary infections as $0.36$ \citep{isoda2025serosurvey}, the MHRW algorithm is preferable as it provides more representative estimates, which is particularly valuable for optimizing intervention policies.

\section{Conclusions and Further Directions}\label{conclusion}
In this study, we assessed the performance of the Random Walk (RW) and Metropolis-Hastings Random Walk (MHRW) sampling algorithms in estimating disease transmission metrics across three network types: \ER~(ER), Small-world (SW), and Scale-Free (SF). We chose these network types because they capture essential properties of real-world networks. For instance, sexual contact networks often resemble ER networks \citep{holme2013extinction}, the SARS-CoV-2 transmission network in Houston, Texas, has been shown to align more closely with SF networks \citep{fujimoto2023beyond}, and the rabies transmission network of Serengeti Lions is observed to be similar to the SW networks \citep{craft2011disease}. Using a stochastic Susceptible-Infected-Recovered (SIR) model and the RW algorithm, we examined how size bias, the over-representation of highly connected individuals in samples, leads to biased estimates of disease spread. Our analysis focused on three key disease metrics: the proportion of infected individuals (epidemic size), the average number of secondary infections (effective reproduction number), and the time-to-infection. Additionally, we investigated how the inclusion of duplicate nodes influences the estimation of key disease metrics.

Our findings in Section \ref{disesase_metrics} demonstrate that the MHRW algorithm is suitable for ER and SW networks. The RW algorithm produced estimates with significant overestimation, approximately $25\%$, for the proportion of infected individuals and secondary infections. In contrast, by reducing size bias, MHRW samples consistently produced estimates of the three disease metrics that closely aligned with the underlying network's estimates, particularly when duplicate sampled nodes were retained. Hence, the RW algorithm is less reliable for precise estimates but can be valuable for early interventions in high-fatality and/or fast-spreading epidemics, such as COVID-19 and Ebola. Conversely, the MHRW algorithm is more suitable for slower and lower-severity epidemics, such as seasonal influenza. Notably, for both sampling algorithms, estimates of time-to-infection align closely with the underlying network for all three network types.

In SF networks, the inherent heterogeneity and presence of high-degree hubs introduced substantial variability in the estimates derived from both the RW and MHRW samples. The high variability in the estimated disease metrics suggests that neither method is optimal for SF networks. Alternative approaches, such as separate sampling of core and peripheral nodes using algorithms like the Sampling for Large-Scale Networks at Low Sampling Rates (SLSR) \citep{Jiao_2024}, may be required for more representative sampling.

In Section \ref{cattle data}, we analysed cattle movement network data that exhibit hybrid characteristics from ER, SW, and SF networks; this further supported the findings above. MHRW sample estimates were more representative of the underlying network when duplicate nodes were retained, while RW samples significantly overestimated the proportion of the infected farms by approximately $100\%$ and the number of secondary infections by over $900\%$ due to size bias. Also, RW samples underestimated the time-to-infection by approximately $40\%$. These findings suggest that RW's conservative estimates are suitable for rapid interventions in highly contagious diseases, such as the Foot-and-Mouth disease. Conversely, the more precise estimates from MHRW sampling are better suited for resource-efficient intervention strategies in less severe and slower epidemics, such as Bovine Viral Diarrhoea.

Our analysis revealed that the sensitivity of both sampling algorithms to recovery rates significantly impacts the accuracy of disease metric estimates (\ref{gammavar}). The difference in the proportion of infected nodes estimated by the two sampling algorithms decreases with an increase in the recovery rate. 

\rev{In this study, we have demonstrated the effectiveness of the MHRW algorithm over the RW algorithm for reducing size bias in estimating disease metrics for three key network types. However, it is essential to note that MHRW, despite its advantages in reducing size bias, has limitations in networks with high-degree variance, where slow mixing can lead to non-representative samples, such as SF networks. Additionally, MHRW relies on the assumption that node degrees are known or accessible during sampling, which may not always be feasible in dynamic networks. While retaining duplicate nodes improves the accuracy of the MHRW estimates, it increases computational costs and may reduce the scalability of the method in larger, more complex networks.}

\rev{The use of the Gillespie algorithm for simulating stochastic SIR disease dynamics also introduces certain constraints; it can become computationally expensive for larger networks, limiting the ability to perform large-scale simulations. Additionally, the choice of parameter values for transmission and recovery rates may not fully capture the diversity of epidemic dynamics in different settings. Although we selected parameters that reflect realistic disease scenarios, the variability in disease dynamics across different populations and network structures may affect the generalizability of our results. To account for variability in network topology, we generated $10,000$ networks of each type, considering specific values for the number of nodes, average degree, edge creation probability, and power-law exponent. However, these network parameters do not cover the full spectrum of all possible topologies within each network family. Consequently, further research is necessary to assess how variations in network structure and epidemiological parameters influence the robustness of our findings.} 

\rev{Despite these limitations, our findings provide valuable insights into the behaviour of the RW and MHRW sampling algorithms and suggest areas for further work. }To further understand the subtle differences between the RW and MHRW sampling algorithms with respect to network structure, it would be valuable to explore the impact of selecting a specific starting node rather than a random node, particularly for SF networks. Furthermore, as observed in Figure \ref{with_dup}, the difference in disease estimates between the RW and MHRW samples decreases with the recovery rate. It would be interesting to determine the value of the recovery rate at which this change begins, as it could guide the choice of sampling algorithm. Additionally, exploring the sensitivity of these estimates to variations in network parameters, such as network size, average degree, and for SF networks, the power-law exponent, would provide further insights.

\rev{To provide deeper insights into the mechanisms driving size bias in sampling, one could study a configuration model with a Negative Binomial (NB) degree distribution, which allows systematic variation of the dispersion parameter to interpolate between Poisson-like (light-tailed) and heavy-tailed (power-law) degree distributions. We explored this approach and found several important subtleties. Notably, when the dispersion parameter $(r)$ is large, for example, $r = 100$, the NB degree distribution closely resembles that of ER networks, characterised by homogeneous connectivity and a few high-degree nodes. This overlap arises because both NB and binomial distributions converge toward a Poisson distribution under specific conditions. On the other hand, decreasing the dispersion parameter increases the variance of the NB degree distribution, yielding heavier tails and greater heterogeneity. However, maintaining an average degree of $5$, as chosen in this study, with low dispersion values (for example, $r = 0.001$ or $0.01$) results in a large number of nodes becoming isolated, significantly reducing the size of the connected component and, consequently, comparability. For example, networks generated using $r = 0.01$ with $10,000$ nodes and average degree $5$ result in fewer than $1,000$ connected nodes, rendering them unsuitable for comparison with the ER, SW, and SF networks used in this study. Thus, although NB-based configuration models offer a promising direction for understanding size bias in relation to network heterogeneity, careful consideration of dispersion effects on network connectivity and comparability is essential. We recommend future work to explore this space further, potentially relaxing constraints on the average degree or network size to investigate size bias under more extreme connectivity profiles.}\\

\noindent\textbf{CRediT authorship contribution statement}\\ \\
\textbf{Neha Bansal:} Conceptualization, Formal analysis, Investigation, Methodology, Validation, Visualization, Writing – original draft \& editing. \textbf{Katerina Kaouri:} Conceptualization, Formal analysis, Investigation, Methodology, Project administration, Supervision, Validation, Writing – review \& editing. \textbf{Thomas E. Woolley:} Conceptualization, Formal analysis, Investigation, Methodology, Project administration, Supervision, Validation, Writing – review \& editing.\\

\noindent\textbf{Declaration of competing interest} \\ \\
The authors declare that they have no known competing financial interests or personal relationships that could have appeared to influence the work reported in this paper.\\

\noindent\textbf{Funding}\\ \\
This work is supported by the Natural Environment, Biotechnology and Biological Sciences and Medical Research councils (NERC, BBSRC and MRC) [grant number: NE/X016714/1] as part of the One Health for One Environment: an A-Z Approach for Tackling Zoonoses (`OneZoo') Centre for Doctoral Training. This work is also supported by an EPSRC Impact Acceleration Account grant, under grant number EP/X525522/1. \\

\noindent\textbf{Data and Code}\\ \\
Data and code are available here: \href{Github}{https://github.com/nehabansal26/Epidemics-on-Networks}

\appendix
\section{Pseudocode for Sampling Algorithms}\label{pseudocode}
We provide the logic of the code for extracting samples using the Random Walk (RW) (Algorithm \ref{alg:RW}) and Metropolis-Hastings Random Walk (MHRW) (Algorithm \ref{alg:MHRW}) algorithms; refer to Section \ref{sample_algo} for the mathematical formulation of the algorithms. For both algorithms, to obtain a sample from a graph, we provide three inputs: i) a graph object, which contains the index or label of nodes and the list of connections between nodes; ii) the number of samples, and iii) the sample size, which is the number of nodes required in a sample.

\begin{algorithm}
\caption{Random Walk (RW) Sampling}\label{alg:RW}
\begin{algorithmic}[1]
\State \textbf{Input:} Graph $G$; Number of samples $W$; Sample size $S$
\State $N \gets$ Number of nodes in $G$  
\State $E \gets$ Edges in $G$
\State Initialize transition matrix $T \gets 0^{N \times N}$
\For{each edge $(i,j)$ in $E$}
    \If{$i \neq j$}
        \State $T[i,j] \gets 1 / \text{degree of node} [i]$
        \State $T[j,i] \gets 1 / \text{degree of node} [j]$
    \EndIf
\EndFor

\State $T\_non\_zero \gets$ matrix of non-zero indices in $T$
\State Initialize sample matrix $sample\_array \gets 0^{W \times S}$
\State $sample\_array[:,0] \gets$ random sample of $W$ nodes from $G$ as seed nodes

\For{$itr = 0$ to $S - 1$}
    \For{each node $i$ in $sample\_array[:,itr]$}
        \State $next\_node \gets$ randomly select a neighbour of node $i$ from $T\_non\_zero[i]$
        \State $sample\_array[i, itr+1] \gets next\_node$
    \EndFor
\EndFor
\end{algorithmic}
\end{algorithm}
\begin{algorithm}
\caption{Metropolis-Hastings Random Walk (MHRW) Sampling}\label{alg:MHRW}
\begin{algorithmic}[1]
\State \textbf{Input:} Graph $G$; Number of samples $W$; Sample size $S$
\State $N \gets$ Number of nodes in $G$  
\State $E \gets$ Edges in $G$
\State Initialize $T \gets 0^{N \times N}$
\For{each edge $(i,j) \in E$}
    \If{$i \neq j$}
        \State $T[i,j] \gets   \min\Big(1, \frac{\text{degree of node} [i]}{\text{degree of node} [j]}\Big)$
        \State $T[j,i] \gets \min\Big(1, \frac{\text{degree of node} [j]}{\text{degree of node} [i]}\Big)$
    \EndIf
\EndFor

\State $random\_probs \gets$ matrix of size $W \times S$ with random numbers from $0$ to $1$
\State Initialize sample matrix $sample\_array \gets 0^{W \times S}$
\State $sample\_array[:,0] \gets$ random sample of $W$ nodes from $G$ as seed nodes

\For{$itr = 0$ to $S - 1$}
    \For{each node $i \in sample\_array[:,itr]$}
        \State $choices[i] \gets$ list of neighbours of node $i$ where $T[i,:] > random\_probs[i,itr]$
        \If{$choices[i]$ not empty}
            \State $next\_node \gets$ randomly select a node from $choices[i]$
        \Else 
            \State $next\_node \gets$  $sample\_array[i,itr]$ 
        \EndIf
        \State $sample\_array[i, itr+1] \gets next\_node$
    \EndFor
\EndFor
\end{algorithmic}
\end{algorithm}

\section{Statistical test results}\label{stat_results}
We discuss the results of the normality test on the data for two disease metrics: proportion of infected nodes and number of secondary infections. We perform the normality test with the hypotheses stated in Section \ref{hyp_all} on the data of disease metrics from the UN, the RW samples, and the MHRW samples. First, we assess whether the data follows a Normal distribution or not, as it influences the choice of statistical tests for comparing the disease metrics among the RW, MHRW, and UN data. If the data is normally distributed, then parametric tests will be used; otherwise, non-parametric tests will be used.

We use the one-sample Kolmogorov-Smirnov (KS) test with a significance level of $\alpha=0.05$ to assess the normality hypotheses \ref{normal_test_pi} and \ref{normal_test_si}. The KS test statistic is the measure of the largest absolute difference between the cumulative distribution of the data (UN, RW, MHRW) and the normal distribution, which ranges from $0$ to $1$. A larger KS statistic indicates a greater deviation from the normal distribution. A KS statistics value closer to $1$ with a p-value $<0.05$ suggests that the data distribution is significantly different from the normal distribution, while a value closer to $0$ with a p-value $>0.05$ implies that the data is significantly similar to a normal distribution. Below are the results for the two disease metrics across three networks (ER, SW, and SF) for the data from the UN, RW, and MHRW samples.
\begin{enumerate}
    \item \textbf{Proportion Infected Nodes:} For all three networks, the KS statistic (Table \ref{infnodes}) values are approximately $0.5$, with p-values $<0.05$, evidence to reject the $H_0$ in Hypothesis \ref{normal_test_pi} for the UN, RW and MHRW data. The KS statistic is sufficiently large, leading us to conclude that the data distributions for the proportion of infected nodes significantly deviate from a Normal distribution. 
    \item \textbf{Secondary Infections:} For all three networks, the KS statistic (Table \ref{secondinfs}) values are approximately $0.5$, with p-values $<0.05$, evidence to reject the $H_0$ in Hypothesis \ref{normal_test_si} for the UN, RW and MHRW data. The KS statistic is sufficiently large, leading us to conclude that the data distributions for the number of secondary infections significantly deviate from a Normal distribution. 
    \item \textbf{Time to Infections:} The KS statistic (Table \ref{time_to_inf}) values are approximately $0.7$ for ER, $0.6$ for SW, and $0.5$ for SF, with p-values $<0.05$, evidence to reject the $H_0$ in Hypothesis \ref{normal_test_time_inf} for the UN, RW and MHRW data. The KS statistic is sufficiently large, leading us to conclude that the data distributions for the number of secondary infections significantly deviate from a Normal distribution. 
\end{enumerate}
\begin{table}[h]
    \centering
    \begin{minipage}{0.45\textwidth}
        \centering
        \begin{tabular}{|c|c|c|c|}
        \hline
        \textbf{Network} & \textbf{UN} & \textbf{RW} & \textbf{MHRW} \\
        \hline
        ER & 0.504 & 0.501 & 0.501 \\
        \hline
        SW & 0.504 & 0.501 & 0.501 \\
        \hline
        SF & 0.504 & 0.502 & 0.507 \\
        \hline
        \end{tabular}
        \caption{One-sample KS test statistics for the proportion of infected nodes (Hypothesis \ref{normal_test_pi}).}
        \label{infnodes}
    \end{minipage}\hfill 
    \begin{minipage}{0.5\textwidth}
        \centering
        \begin{tabular}{|c|c|c|c|}
        \hline
        \textbf{Network} & \textbf{UN} & \textbf{RW} & \textbf{MHRW} \\
        \hline
        ER & 0.578 & 0.599 & 0.582 \\
        \hline
        SW & 0.509 & 0.500 & 0.500 \\
        \hline
        SF & 0.509 & 0.500& 0.500 \\
        \hline
        \end{tabular}
        \caption{One-sample KS test statistics for the number of secondary infections (Hypothesis \ref{normal_test_si}).}
        \label{secondinfs}
    \end{minipage}
\end{table}
\begin{table}[h]
    \centering
\begin{tabular}{|c|c|c|c|}
        \hline
        \textbf{Network} & \textbf{UN} & \textbf{RW} & \textbf{MHRW} \\
        \hline
        ER & 0.729 & 0.721 & 0.726 \\
        \hline
        SW & 0.659 & 0.663 & 0.662 \\
        \hline
        SF & 0.500 & 0.500& 0.500 \\
        \hline
\end{tabular}
\caption{One-sample KS test statistics for the number of secondary infections (Hypothesis \ref{normal_test_time_inf}).}
    \label{time_to_inf}
\end{table}
Since, data does not follow a Normal distribution across networks, for all three disease metrics, a non-parametric statistical test is required to compare the RW and MHRW samples with the UN. We are using the Man-Whitney U-test in this work; results for the same are discussed in Section \ref{results}.

\section{Disease metrics}
We compare the sampling algorithms (RW and MHRW) based on the accuracy of disease metric estimates for the ER, SW and SF networks using the method described in Section \ref{sim_s}. Specifically, in Section \ref{gammavar}, we compare the sampling algorithms for three disease metrics: 1) proportion of infected nodes, 2) average number of secondary infections, and 3) time-to-infection for the ER and SW networks, with $\beta$ values from $0$ to $1$ and $\gamma \in \{1/7,1/14\}$.

\subsection{Sensitivity analysis - Recovery rate}\label{gammavar}
In Section \ref{results}, we discussed the results obtained with a varying transmission rate $\beta$ and a fixed $\ ,gamma=1$. Here, we investigate the impact of varying $\gamma$ along with $\beta$ on the accuracy of disease estimates from sampling algorithms for the ER and SW networks. We compare the disease metric estimates from both sampling algorithms (RW and MHRW) with the UN, using three values of the recovery rate $\gamma$: $1$, $1/7$, and $1/14$. We also discuss the results based on two scenarios: with duplicate nodes in the samples and without duplicate nodes in the samples.

\subsubsection{Samples with duplicate nodes}
\begin{figure}
    \centering
    \begin{minipage}{\textwidth}
        \centering
        \caption*{(1) Proportion of infected nodes.}
        \subfig{0.23}{1}{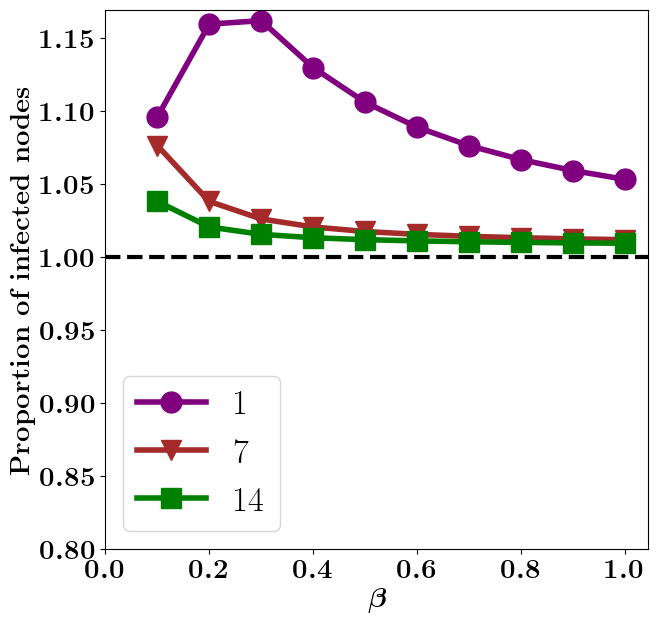}{\ER; RW}
        \subfig{0.23}{1}{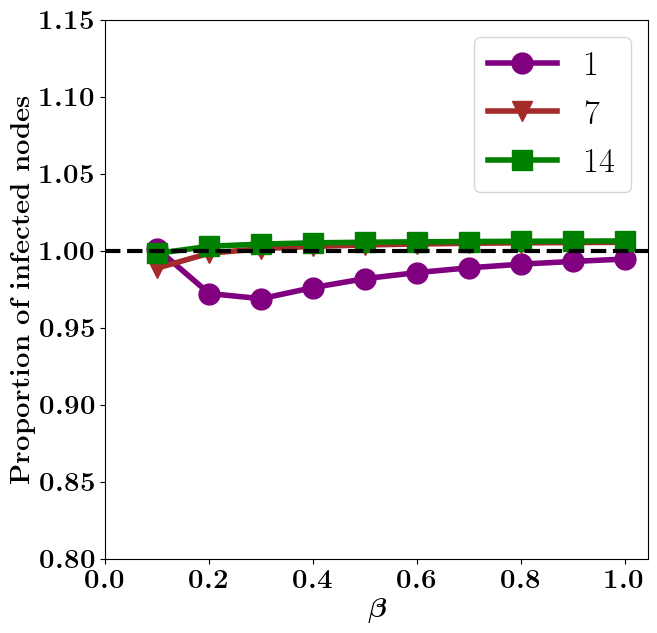}{\ER; MHRW}
        \subfig{0.23}{1}{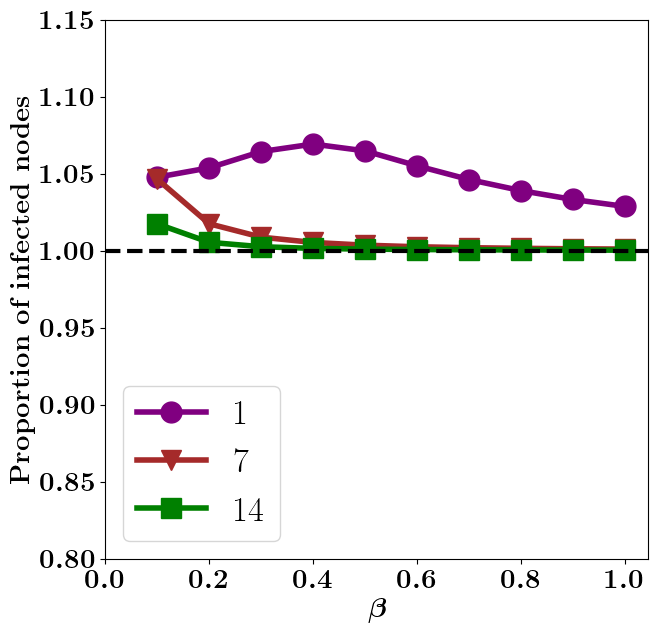}{Small-world; RW}
        \subfig{0.23}{1}{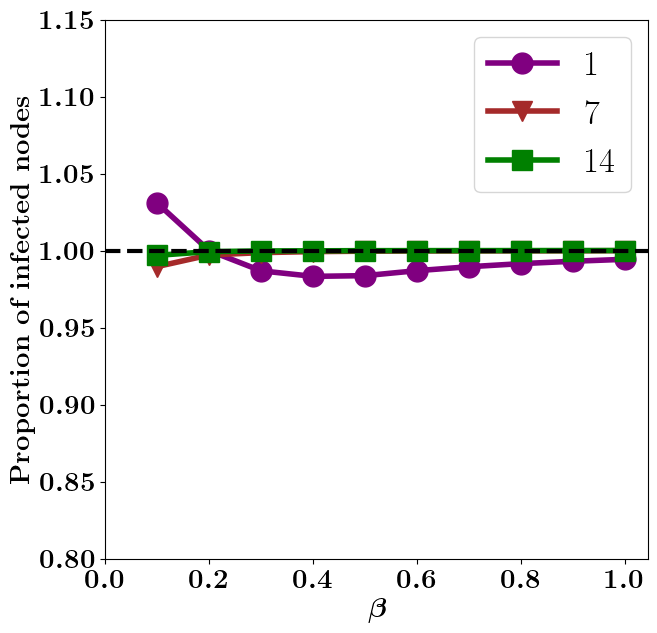}{Small-world; MHRW}
    \end{minipage}
    \begin{minipage}\textwidth
        \centering
        \caption*{(2) Number of secondary infections.}
        \subfig{0.23}{1}{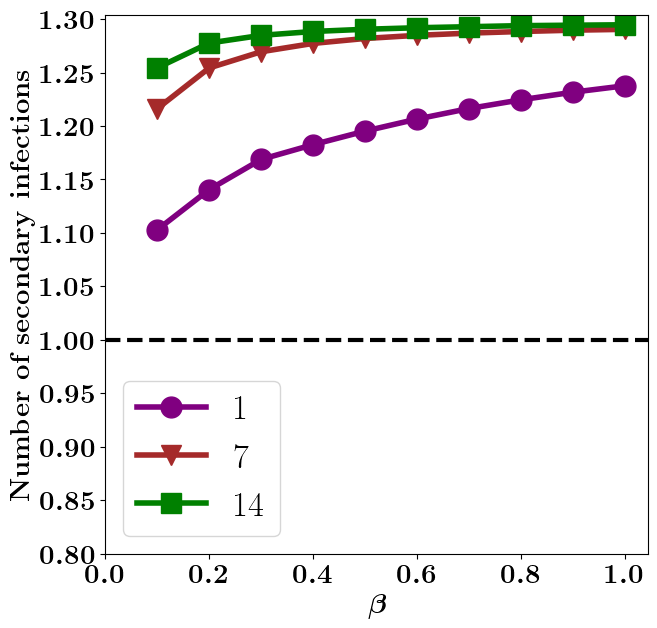}{\ER; RW}
        \subfig{0.23}{1}{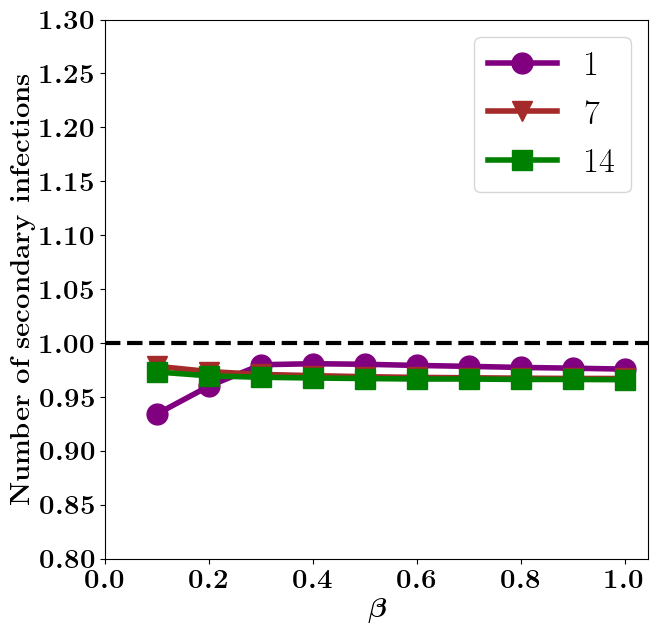}{\ER; MHRW}
        \subfig{0.23}{1}{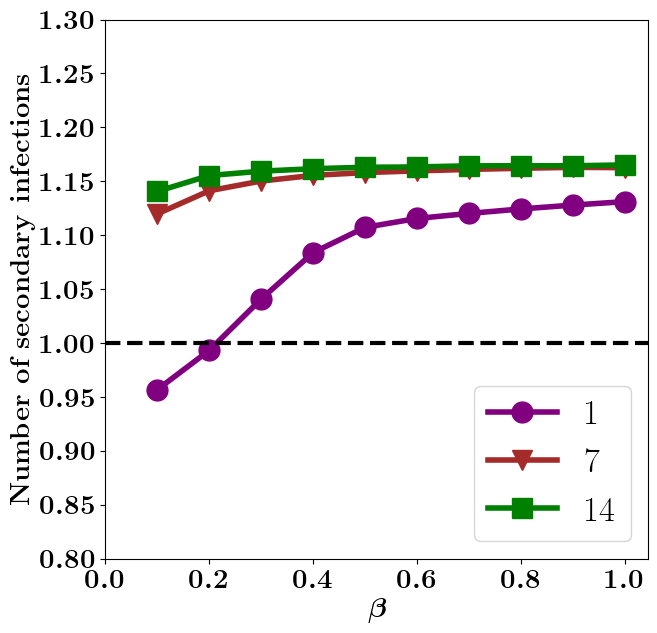}{Small-world; RW}
        \subfig{0.23}{1}{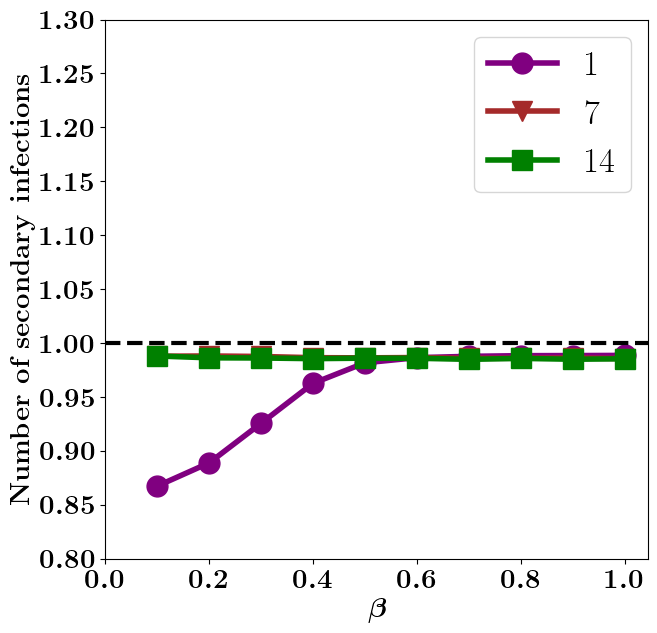}{Small-world; MHRW}
    \end{minipage}
    \begin{minipage}{\textwidth}
        \centering
        \caption*{(3) time-to-infection (days).}
        \subfig{0.23}{1}{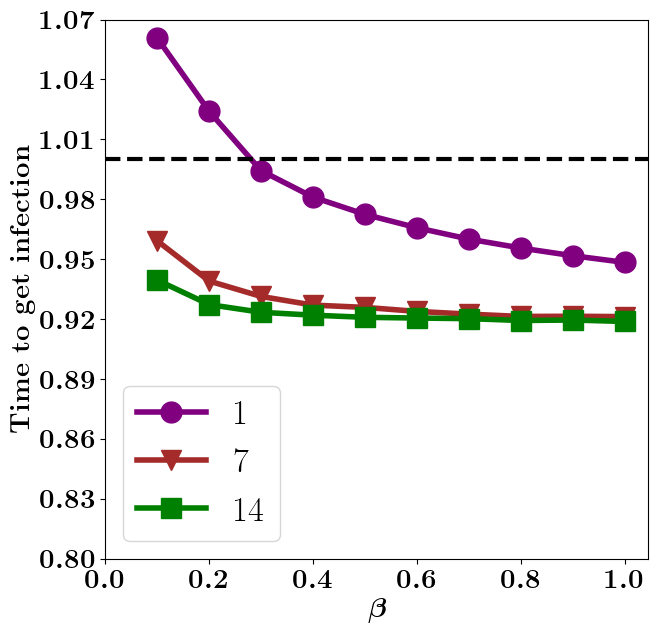}{\ER; RW}
        \subfig{0.23}{1}{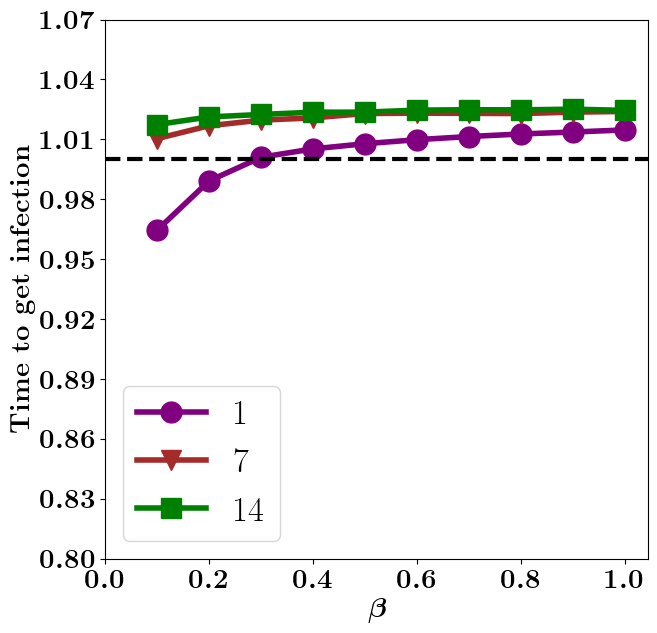}{\ER; MHRW}
        \subfig{0.23}{1}{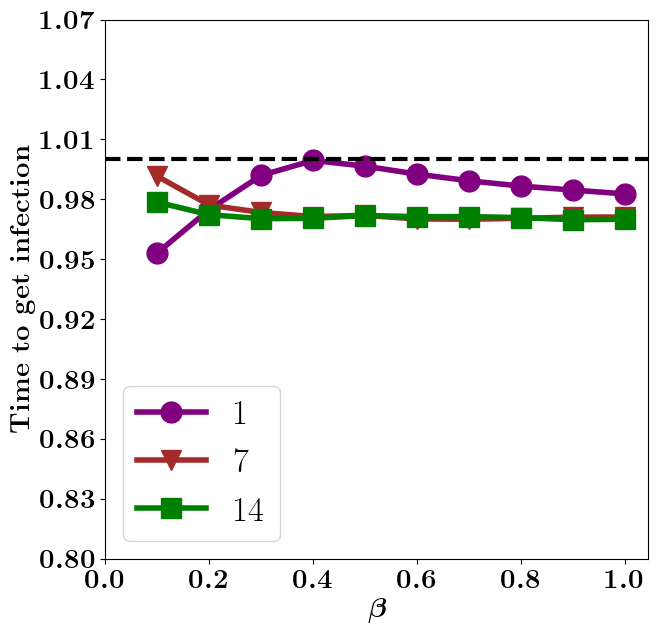}{Small-world; RW}
        \subfig{0.23}{1}{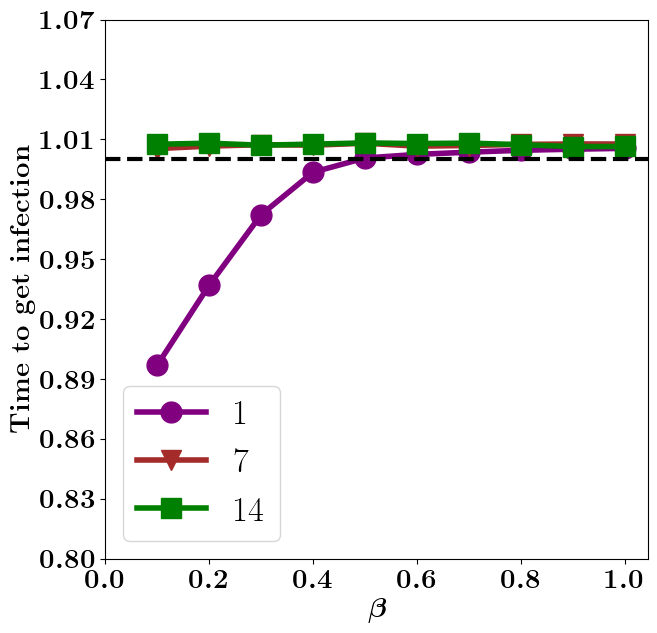}{Small-world; MHRW}
    \end{minipage}
    \caption{The average value of the ratio of three disease metrics estimates for ER and SW networks using RW and MHRW samples (with duplicate nodes) relative to the UN; for varying recovery rate $\gamma \in \{1,1/7,1/14\}$ and transmission rate $\beta$ from $0$ to $1$. The UN estimates are based on SIR model simulations for $10,000$ networks of each type (the parameters are found in Table \ref{net_params_sim}) generated with the NetworkX library \citep{hagberg2008exploring}. Sample estimates are derived from $100$ samples (size $500$) for $10,000$ networks using RW and MHRW.}
\label{with_dup}
\end{figure}

In Figure \ref{with_dup}, we compare the average ratio of disease metric estimates between the RW and MHRW samples (with duplicate nodes) with the UN for three disease metrics: i) proportion of infected nodes, ii) number of secondary infections, and iii) time to get infected in days. The recovery rate ($\gamma$) values are $1,1/7$, and $1/14$ and transmission rate $\beta$ increases from $0$ to $1$. The purple line plots with circle markers are for $\gamma=1$, the brown line plots with triangle markers are for $\gamma=1/7$, and the green line plots with square markers are for $\gamma=1/14$. For a network and a disease metric, there are two plots, one for the RW sample estimates and another for the MHRW sample estimates.

For the ER network, as shown in Figures \ref{node_idx_mean_ER_RW} and \ref{node_idx_mean_ER_MHRW}, the sample estimates for the proportion of infected nodes become closer to the UN as the recovery rate decreases from $1$ to $1/14$, for both sampling algorithms. Especially, the MHRW sample estimates are well aligned with the UN for $\gamma = 1/14$. For the RW sample estimates, there is a significant drop in overestimation with the decrease in the recovery rate from $1$ to $1/14$. We observe similar behaviour for the SW networks, as shown in Figures \ref{node_idx_mean_SW_RW} and \ref{node_idx_mean_SW_MHRW}. 

For the ER network, as shown in Figures \ref{second_inf_mean_ER_RW} and \ref{second_inf_mean_ER_MHRW}, the sample estimates for the number of secondary infections deviate far from the UN as the recovery rate decreases from $1$ to $1/14$, for both sampling algorithms. The MHRW sample estimates are closer to the UN compared to the RW sample estimates, irrespective of the recovery rate values. We observe similar behaviour for the SW networks, as shown in Figures \ref{second_inf_mean_SW_RW} and \ref{second_inf_mean_SW_MHRW}. 

For the estimates of the time-to-infection in the ER network, as shown in Figures \ref{inf_time_mean_ER_RW} and \ref{inf_time_mean_ER_MHRW}, the sample estimates deviate away from the UN for both sampling algorithms as recovery rate decreases. The overestimation increases for the MHRW samples, and the underestimation increases for the RW sample estimates. For the SW networks, as shown in Figures \ref{inf_time_mean_SW_RW} and \ref{inf_time_mean_SW_MHRW}, underestimation increases for the RW sample estimates. At the same time, the MHRW sample estimates align well with the UN, showing a decrease in the recovery rate from $1$ to $1/14$ for $\beta < 0.4$. 
\subsubsection{Samples without duplicate nodes}
\begin{figure}
    \centering
    \begin{minipage}{\textwidth}
        \centering
        \caption*{(1) Proportion of infected nodes.}
        \subfig{0.23}{1}{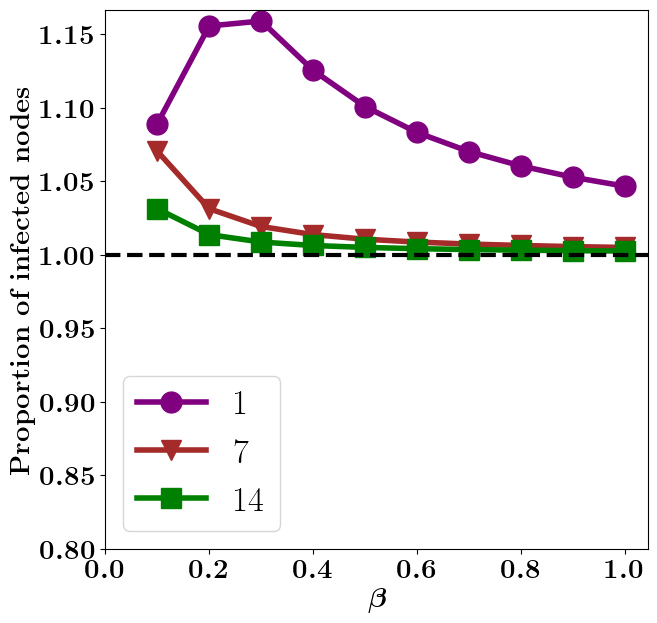}{\ER; RW}
        \subfig{0.23}{1}{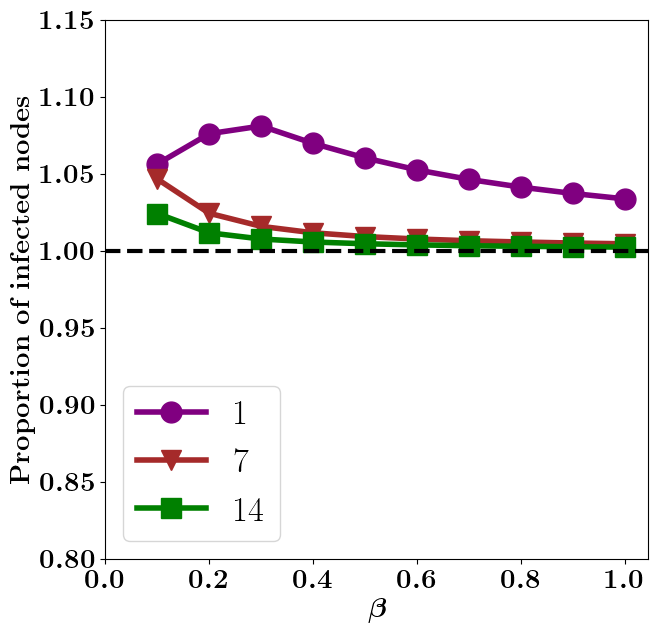}{\ER; MHRW}
        \subfig{0.23}{1}{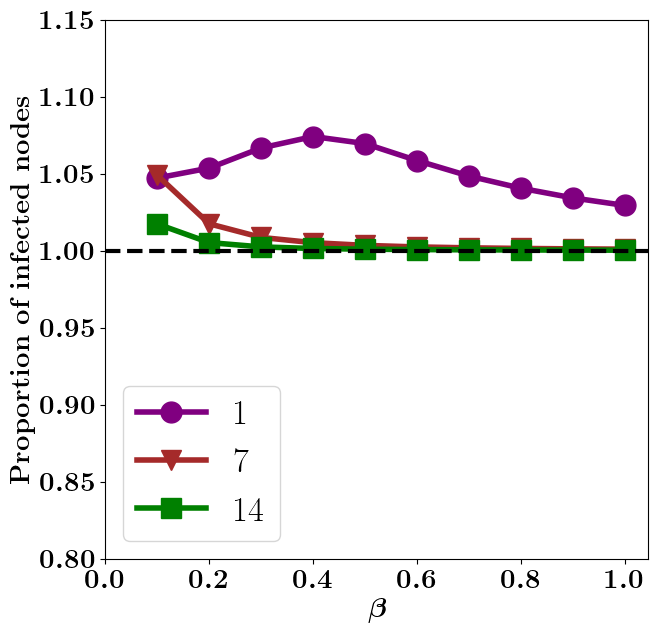}{Small-world; RW}
        \subfig{0.23}{1}{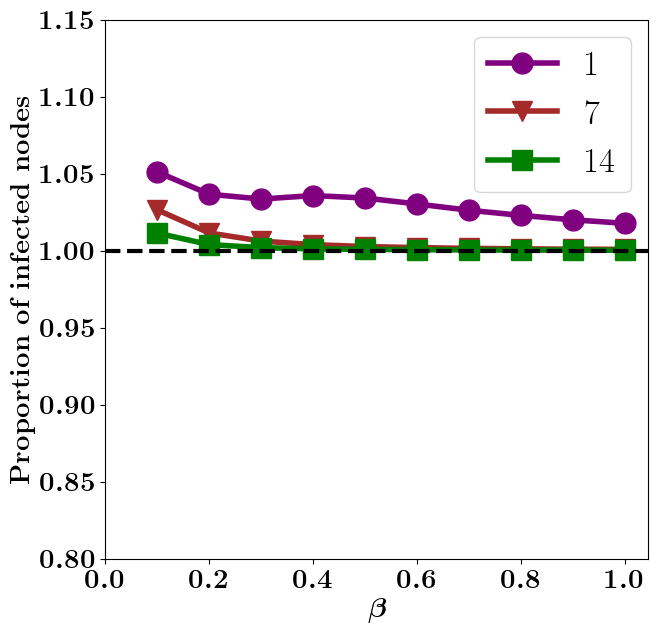}{Small-world; MHRW}
    \end{minipage}
    \begin{minipage}\textwidth
        \centering
        \caption*{(2) Number of secondary infections.}
        \subfig{0.23}{1}{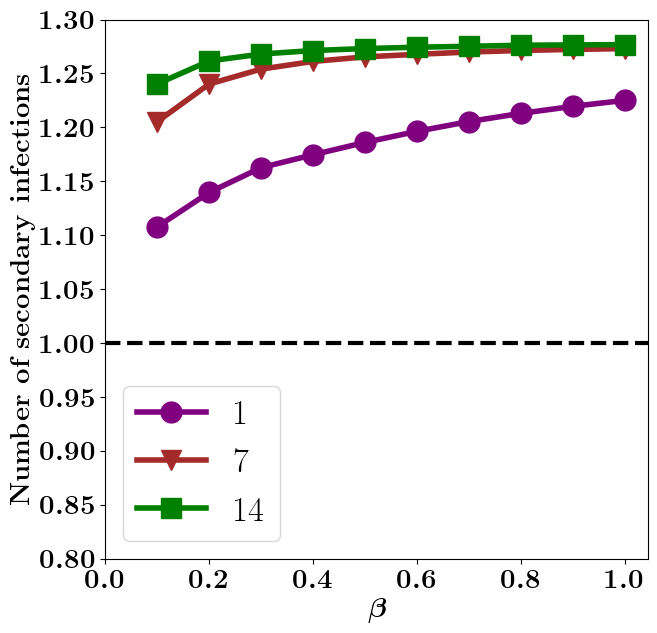}{\ER; RW}
        \subfig{0.23}{1}{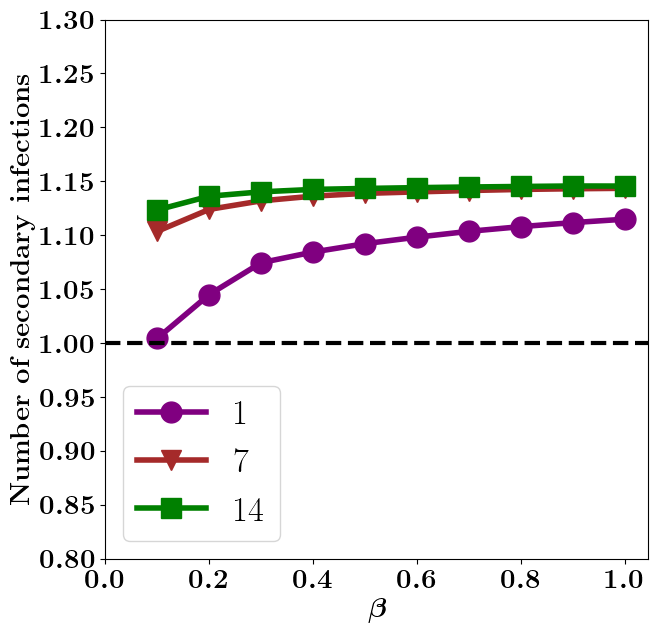}{\ER; MHRW}
        \subfig{0.23}{1}{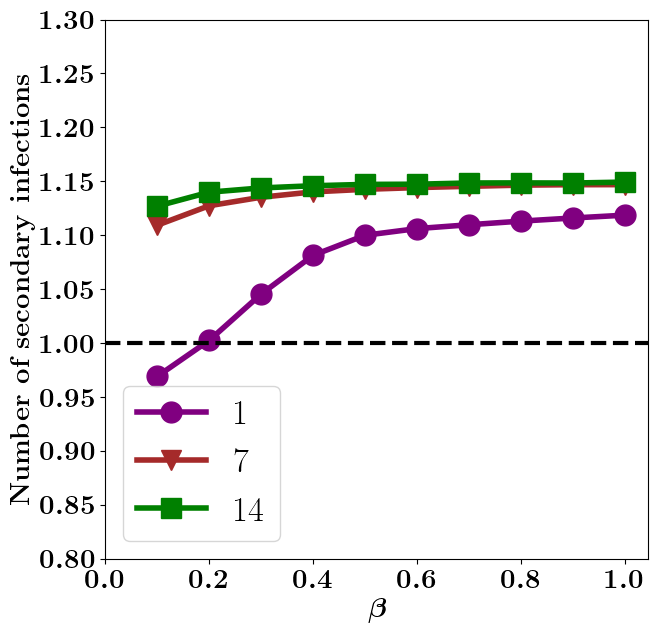}{Small-world; RW}
        \subfig{0.23}{1}{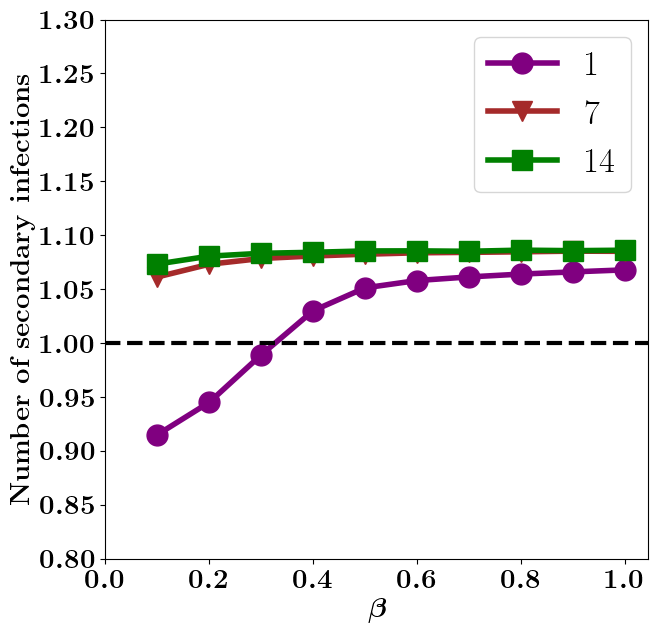}{Small-world; MHRW}
    \end{minipage}
    \begin{minipage}{\textwidth}
        \centering
        \caption*{(3) time-to-infection (days).}
        \subfig{0.23}{1}{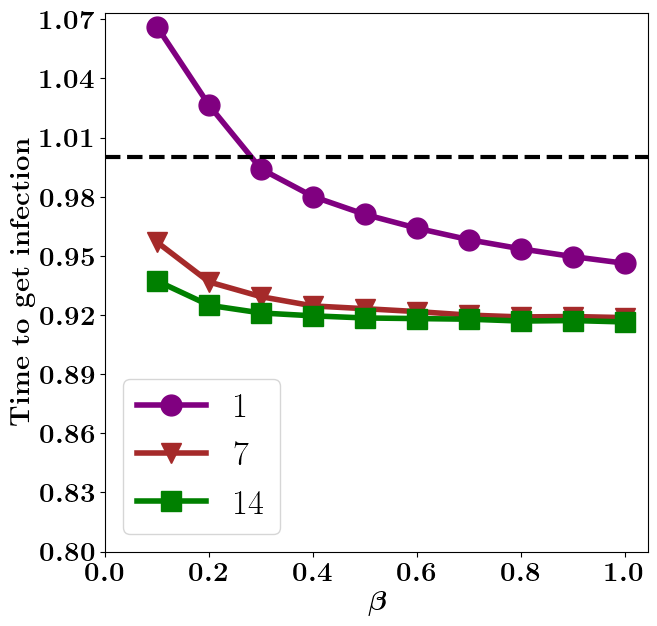}{\ER; RW}
        \subfig{0.23}{1}{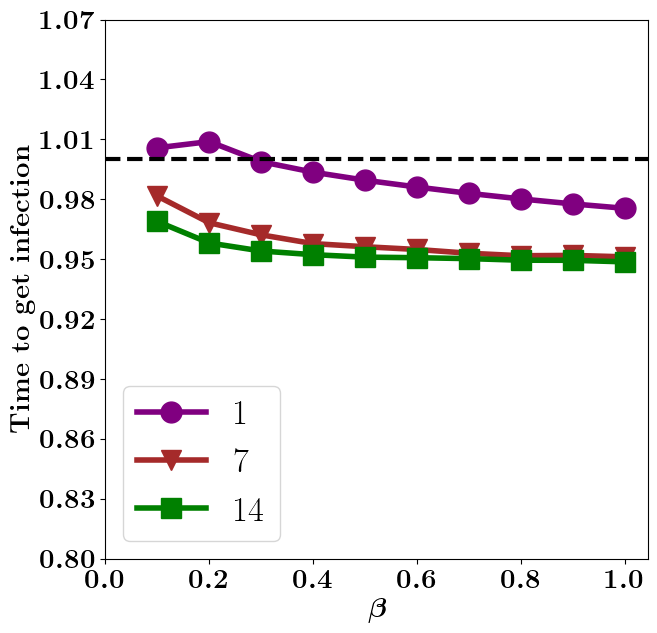}{\ER; MHRW}
        \subfig{0.23}{1}{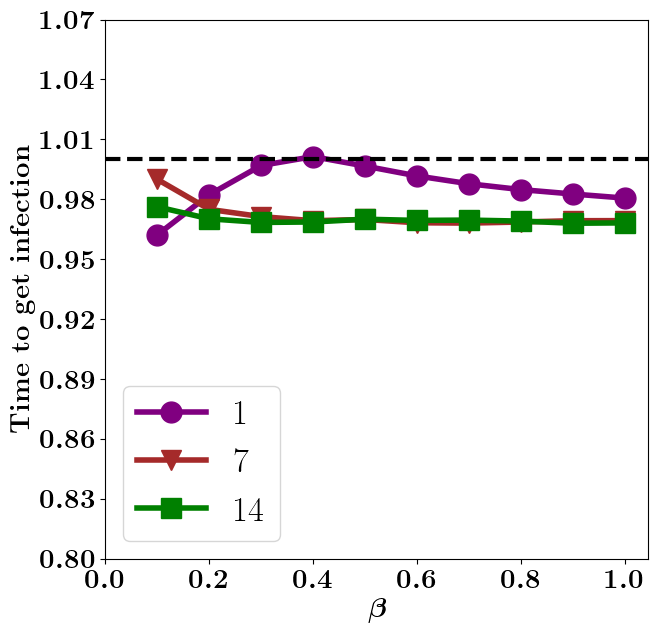}{Small-world; RW}
        \subfig{0.23}{1}{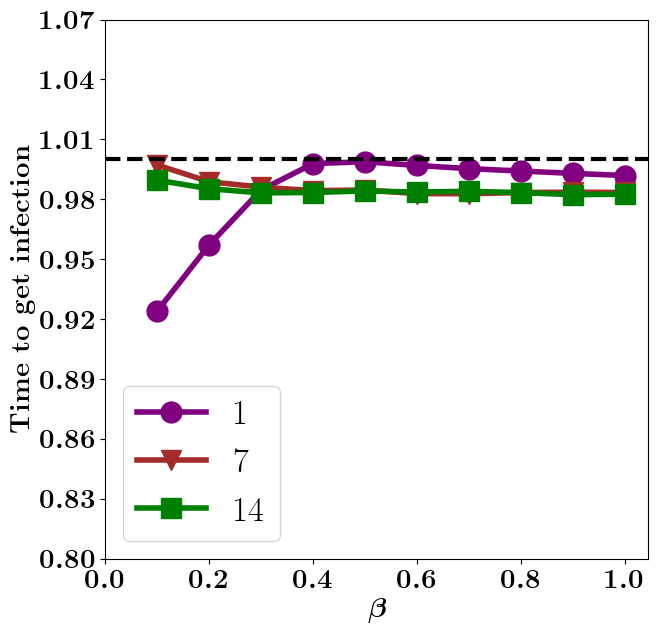}{Small-world; MHRW}
    \end{minipage}
    \caption{The ratio of the average value of three disease metrics estimates for ER and SW networks using RW and MHRW samples (without duplicate nodes) relative to the UN; for varying recovery rate $\gamma \in \{1,1/7,1/14\}$ and transmission rate $\beta$ from $0$ to $1$.The UN estimates are based on SIR model simulations ($\gamma=1$) for $10,000$ networks of each type (the parameters are found in Table \ref{net_params_sim}) generated with the NetworkX library \citep{hagberg2008exploring}. Sample estimates are derived from $100$ samples (size $500$) for $10,000$ networks using RW and MHRW.}
    \label{without_dup}
\end{figure}

In Figure \ref{without_dup}, we compare the average ratio of disease metric estimates between the RW and MHRW samples (without duplicate nodes) with the UN for three disease metrics: i) proportion of infected nodes, ii) number of secondary infections, and iii) time to get infected in days. The recovery rate ($\gamma$) values are $1,1/7$, and $1/14$, and transmission rate $\beta$ increases from $0$ to $1$. The purple line plots with circle markers are for $\gamma=1$, the brown line plots with triangle markers are for $\gamma=1/7$, and the green line plots with square markers are for $\gamma=1/14$. For a network and a disease metric, there are two plots, one for the RW sample estimates and another for the MHRW sample estimates. 

For the ER network, as shown in Figures \ref{node_idx_mean_ER_RW_dup} and \ref{node_idx_mean_ER_MHRW_dup}, the sample estimates for the proportion of infected nodes become closer to the UN as the recovery rate decreases from $1$ to $1/14$, for both sampling algorithms. For the RW sample estimates, we observe a significant drop in overestimation with the decrease in recovery rate from $1$ to $1/14$. We observe similar behaviour for the SW networks, as shown in Figures \ref{node_idx_mean_SW_RW_dup} and \ref{node_idx_mean_SW_MHRW_dup}. 

For the ER network, as shown in Figures \ref{second_inf_mean_ER_RW_dup} and \ref{second_inf_mean_ER_MHRW_dup}, the sample estimates for the number of secondary infections deviates far from the UN as the recovery rate decreases from $1$ to $1/14$, for both sampling algorithms. We observe similar behaviour for the SW networks, as shown in Figures \ref{second_inf_mean_SW_RW_dup} and \ref{second_inf_mean_SW_MHRW_dup}. 

For the estimates of the time-to-infection in the ER network, as shown in Figures \ref{inf_time_mean_ER_RW_dup} and \ref{inf_time_mean_ER_MHRW_dup}, we observe that sample estimates deviate away from the UN for both sampling algorithms as the recovery rate decreases. The underestimation increases for the RW sample estimates, and the overestimation increases for the MHRW sample estimates. For the SW networks, as shown in Figures \ref{inf_time_mean_SW_RW_dup} and \ref{inf_time_mean_SW_MHRW_dup}, underestimation increases for the RW sample estimates. At the same time, the MHRW sample estimates align well with the UN, showing a decrease in the recovery rate from $1$ to $1/14$ for $\beta < 0.4$. 

Overall, from comparing the sample estimates for three $\gamma$ values ($1$, $1/7$, and $1/14$), we observe that the estimates for both sampling algorithms are very similar when $\gamma = 1/7$ and $\gamma = 1/14$. However, with $\gamma = 1$, there is a noticeable deviation in the sample estimates. This suggests that the sample estimates are sensitive to both the recovery rate and the type of disease metric, as well as the network type.

 \bibliography{cas-refs}

\begin{thebibliography}{76}
\providecommand{\natexlab}[1]{#1}
\providecommand{\url}[1]{\texttt{#1}}
\expandafter\ifx\csname urlstyle\endcsname\relax
  \providecommand{\doi}[1]{doi: #1}\else
  \providecommand{\doi}{doi: \begingroup \urlstyle{rm}\Url}\fi

\bibitem[Aldous and Wilson(2003)]{aldous2003graphs}
Joan~M Aldous and Robin~J Wilson.
\newblock \emph{Graphs and applications: an introductory approach}.
\newblock Springer Science \& Business Media, 2003.

\bibitem[Arratia et~al.(2019)Arratia, Goldstein, and Kochman]{arratia2019size}
R.~Arratia, L.~Goldstein, and F.~Kochman.
\newblock Size bias for one and all.
\newblock \emph{Probab. Surveys}, 2019.

\bibitem[Avraam and Hadjichrysanthou(2025)]{avraam2025impact}
Demetris Avraam and Christoforos Hadjichrysanthou.
\newblock The impact of contact-network structure on important epidemiological quantities of infectious disease transmission and the identification of the extremes.
\newblock \emph{Journal of Theoretical Biology}, 2025.

\bibitem[Balakrishnan(2019)]{balakrishnan2019exponential}
K.~Balakrishnan.
\newblock Exponential distribution: theory, methods and applications.
\newblock \emph{Routledge}, 2019.

\bibitem[Banerjee and Chaudhury(2010)]{banerjee2010statistics}
A.~Banerjee and S.~Chaudhury.
\newblock Statistics without tears: Populations and samples.
\newblock \emph{Industrial Psychiatry Journal}, 2010.

\bibitem[Bansal et~al.(2010)Bansal, Read, Pourbohloul, and Meyers]{bansal2010dynamic}
S.~Bansal, J.~Read, B.~Pourbohloul, and L.~A. Meyers.
\newblock The dynamic nature of contact networks in infectious disease epidemiology.
\newblock \emph{Journal of Biological Dynamics}, 2010.

\bibitem[Barab{\'a}si and Bonabeau(2003)]{barabasi2003scale}
Albert~L{\'a}szl{\'o} Barab{\'a}si and Eric Bonabeau.
\newblock Scale-free networks.
\newblock \emph{Scientific american}, 2003.

\bibitem[Barabási and Albert(1999)]{barabasi1999emergence}
A.~L. Barabási and R.~Albert.
\newblock Emergence of scaling in random networks.
\newblock \emph{Science}, 1999.

\bibitem[Bekara et~al.(2014)Bekara, Courcoul, Benet, and Durand]{bekara2014modeling}
M.~E. Bekara, A.~Courcoul, J.~Benet, and B.~Durand.
\newblock Modeling tuberculosis dynamics, detection and control in cattle herds.
\newblock \emph{PloS one}, 2014.

\bibitem[Bishop and Nasrabadi(2006)]{bishop2006pattern}
Christopher~M Bishop and Nasser~M Nasrabadi.
\newblock \emph{Pattern recognition and machine learning}.
\newblock Springer, 2006.

\bibitem[Bondy and Zlot(1976)]{bondy1976standard}
W.~H. Bondy and W.~Zlot.
\newblock The standard error of the mean and the difference between means for finite populations.
\newblock \emph{The American Statistician}, 1976.

\bibitem[Brauer(2008)]{Brauer2008}
Fred Brauer.
\newblock \emph{Compartmental Models in Epidemiology}, chapter~2.
\newblock Springer Berlin Heidelberg, 2008.

\bibitem[Chowell et~al.(2008)Chowell, Miller, and Viboud]{chowell2008seasonal}
G.~M. A.~M. Chowell, M.~A. Miller, and C.~Viboud.
\newblock Seasonal influenza in the united states, france, and australia: transmission and prospects for control.
\newblock \emph{Epidemiology \& Infection}, 2008.

\bibitem[Christley et~al.(2005)Christley, Robinson, Lysons, and French]{christley2005network}
R.~M. Christley, S.~E. Robinson, R.~Lysons, and N.~P. French.
\newblock Network analysis of cattle movement in great britain.
\newblock \emph{Proc. Soc. Vet. Epidemiol. Prev. Med}, 2005.

\bibitem[Conroy et~al.(2016)]{conroy2016rcsi}
R.~M. Conroy et~al.
\newblock The rcsi sample size handbook.
\newblock \emph{A Rough Guide}, 2016.

\bibitem[Craft(2015)]{craft2015infectious}
M.~E. Craft.
\newblock Infectious disease transmission and contact networks in wildlife and livestock.
\newblock \emph{Philosophical Transactions of the Royal Society B: Biological Sciences}, 2015.

\bibitem[Craft and Caillaud(2011)]{craft2011network}
M.~E. Craft and D.~Caillaud.
\newblock Network models: an underutilized tool in wildlife epidemiology?
\newblock \emph{Interdisciplinary Perspectives on Infectious Diseases}, 2011.

\bibitem[Craft et~al.(2011)Craft, Volz, Packer, and Meyers]{craft2011disease}
Meggan~E Craft, Erik Volz, Craig Packer, and Lauren~Ancel Meyers.
\newblock Disease transmission in territorial populations: the small-world network of serengeti lions.
\newblock \emph{Journal of the Royal Society Interface}, 2011.

\bibitem[Csefalvay(2023)]{von2023computational}
C.~Von Csefalvay.
\newblock Computational modeling of infectious disease: with applications in python.
\newblock \emph{Elsevier}, 2023.

\bibitem[Cui et~al.(2022)Cui, Li, Li, Wang, and Chen]{cui2022survey}
Y.~Cui, X.~Li, J.~Li, H.~Wang, and X.~Chen.
\newblock A survey of sampling method for social media embeddedness relationship.
\newblock \emph{ACM Computing Surveys}, 2022.

\bibitem[Danon et~al.(2011)Danon, Ford, House, Jewell, Keeling, Roberts, Ross, and Vernon]{danon2011networks}
L.~Danon, A.~P. Ford, T.~House, C.~P. Jewell, M.~J. Keeling, G.~O. Roberts, J.~V. Ross, and M.~C. Vernon.
\newblock Networks and the epidemiology of infectious disease.
\newblock \emph{Interdisciplinary Perspectives on Infectious Diseases}, 2011.

\bibitem[Dempsey et~al.(2012)Dempsey, Duraisamy, Bhowmick, and Ali]{dempsey2012development}
K.~Dempsey, K.~Duraisamy, S.~Bhowmick, and H.~Ali.
\newblock The development of parallel adaptive sampling algorithms for analyzing biological networks.
\newblock In \emph{2012 IEEE 26th International Parallel and Distributed Processing Symposium Workshops \& PhD Forum}. IEEE, 2012.

\bibitem[Dobos et~al.(2024)Dobos, v.~Dobos, and Kiss]{dobos2024control}
A.~Dobos, v.~Dobos, and I.~Kiss.
\newblock How control and eradication of bvdv at farm level influences the occurrence of calf diseases and antimicrobial usage during the first six months of calf rearing.
\newblock \emph{Irish Veterinary Journal}, 2024.

\bibitem[Duncan et~al.(2022)Duncan, Reeves, Gunn, and Humphry]{duncan2022quantifying}
A.~J. Duncan, A.~Reeves, G.~J. Gunn, and R.~W. Humphry.
\newblock Quantifying changes in the british cattle movement network.
\newblock \emph{Preventive Veterinary Medicine}, 2022.

\bibitem[Erdős and Rényi(1959)]{erdios1959random}
P.~Erdős and A.~Rényi.
\newblock On random graphs.
\newblock \emph{Publicationes Mathematicae}, 1959.

\bibitem[Erdős et~al.(1960)Erdős, Rényi, et~al.]{erdos1960evolution}
P.~Erdős, A.~Rényi, et~al.
\newblock On the evolution of random graphs.
\newblock \emph{Publ. Math. Inst. Hung. Acad. Sci.}, 1960.

\bibitem[Fielding et~al.(2019)Fielding, McKinley, Silk, Delahay, and McDonald]{fielding2019contact}
H.~R. Fielding, T.~J. McKinley, M.~J. Silk, R.~J. Delahay, and R.~A. McDonald.
\newblock Contact chains of cattle farms in great britain.
\newblock \emph{Royal Society Open Science}, 2019.

\bibitem[Fournet and Barrat(2017)]{fournet2017estimating}
J.~Fournet and A.~Barrat.
\newblock Estimating the epidemic risk using non-uniformly sampled contact data.
\newblock \emph{Scientific Reports}, 2017.

\bibitem[Fujimoto et~al.(2023)Fujimoto, Kuo, Stott, Lewis, Chan, Lyu, Veytsel, Carr, Broussard, Short, et~al.]{fujimoto2023beyond}
Kayo Fujimoto, Jacky Kuo, Guppy Stott, Ryan Lewis, Hei~Kit Chan, Leke Lyu, Gabriella Veytsel, Michelle Carr, Tristan Broussard, Kirstin Short, et~al.
\newblock Beyond scale-free networks: integrating multilayer social networks with molecular clusters in the local spread of covid-19.
\newblock \emph{Scientific Reports}, 2023.

\bibitem[Gillespie(1992)]{gillespie1992rigorous}
D.~T. Gillespie.
\newblock A rigorous derivation of the chemical master equation.
\newblock \emph{Physica A: Statistical Mechanics and its Applications}, 1992.

\bibitem[Gillespie(2007)]{Gillespie_2007}
D.~T. Gillespie.
\newblock Stochastic simulation of chemical kinetics.
\newblock \emph{Annual Review of Physical Chemistry}, 2007.

\bibitem[Gjoka et~al.(2010)Gjoka, Kurant, Butts, and Markopoulou]{gjoka2010walking}
M.~Gjoka, M.~Kurant, C.~T. Butts, and A.~Markopoulou.
\newblock Walking in facebook: a case study of unbiased sampling of osns.
\newblock In \emph{2010 Proceedings IEEE Infocom}. IEEE, 2010.

\bibitem[G{\"o}bel and Jagers(1974)]{gobel1974random}
F.~G{\"o}bel and A.~A. Jagers.
\newblock Random walks on graphs.
\newblock \emph{Stochastic Processes and Their Applications}, 1974.

\bibitem[Hagberg et~al.(2008)Hagberg, Swart, and Schult]{hagberg2008exploring}
A.~Hagberg, P.~J. Swart, and D.~A. Schult.
\newblock Exploring network structure, dynamics, and function using networkx.
\newblock Technical report, Los Alamos National Laboratory (LANL), Los Alamos, NM (United States), 2008.

\bibitem[Harris et~al.(2020)Harris, Millman, van~der Walt, Gommers, Virtanen, Cournapeau, Wieser, Taylor, Berg, Smith, Kern, Picus, Hoyer, van Kerkwijk, Brett, Haldane, del Río, Wiebe, Peterson, Gérard-Marchant, Sheppard, Reddy, Weckesser, Abbasi, Gohlke, and Oliphant]{2020NumPy-Array}
C.~R. Harris, K.~J. Millman, S.~J. van~der Walt, R.~Gommers, P.~Virtanen, D.~Cournapeau, E.~Wieser, J.~Taylor, S.~Berg, N.~J. Smith, R.~Kern, M.~Picus, S.~Hoyer, M.~H. van Kerkwijk, M.~Brett, A.~Haldane, J.~F. del Río, M.~Wiebe, P.~Peterson, P.~Gérard-Marchant, K.~Sheppard, T.~Reddy, W.~Weckesser, H.~Abbasi, C.~Gohlke, and T.~E. Oliphant.
\newblock Array programming with numpy.
\newblock \emph{Nature}, 2020.

\bibitem[Holme(2013)]{holme2013extinction}
Petter Holme.
\newblock Extinction times of epidemic outbreaks in networks.
\newblock \emph{PloS one}, 2013.

\bibitem[Hu and Lau(2013)]{hu2013survey}
P.~Hu and W.~C. Lau.
\newblock A survey and taxonomy of graph sampling.
\newblock \emph{arXiv preprint arXiv:1308.5865}, 2013.

\bibitem[Isoda et~al.(2025)Isoda, Sekiguchi, Ryu, Notsu, Kobayashi, Hamaguchi, Hiono, Ushitani, and Sakoda]{isoda2025serosurvey}
Norikazu Isoda, Satoshi Sekiguchi, Chika Ryu, Kosuke Notsu, Maya Kobayashi, Karin Hamaguchi, Takahiro Hiono, Yuichi Ushitani, and Yoshihiro Sakoda.
\newblock Serosurvey of bovine viral diarrhea virus in cattle in southern japan and estimation of its transmissibility by transient infection in nonvaccinated cattle.
\newblock \emph{Viruses}, 2025.

\bibitem[Jiao(2024)]{Jiao_2024}
Bo~Jiao.
\newblock Sampling unknown large networks restricted by low sampling rates.
\newblock \emph{Scientific Reports}, 2024.

\bibitem[Joyal-Desmarais et~al.(2022)Joyal-Desmarais, Stojanovic, Kennedy, Enticott, Boucher, Vo, Ko{\v{s}}ir, Lavoie, and Bacon]{joyal2022well}
Keven Joyal-Desmarais, Jovana Stojanovic, Eric~B Kennedy, Joanne~C Enticott, Vincent~Gosselin Boucher, Hung Vo, Ur{\v{s}}ka Ko{\v{s}}ir, Kim~L Lavoie, and Simon~L Bacon.
\newblock How well do covariates perform when adjusting for sampling bias in online covid-19 research? insights from multiverse analyses.
\newblock \emph{European Journal of Epidemiology}, 2022.

\bibitem[Jr(1951)]{massey1951kolmogorov}
F.~J.~Massey Jr.
\newblock The kolmogorov-smirnov test for goodness of fit.
\newblock \emph{Journal of the American Statistical Association}, 1951.

\bibitem[Kermack and McKendrick(1927)]{kermack1927contribution}
W.~O. Kermack and A.~G. McKendrick.
\newblock A contribution to the mathematical theory of epidemics.
\newblock \emph{Proceedings of the Royal Society of London. Series A, Containing Papers of a Mathematical and Physical Character}, 1927.

\bibitem[Kiss et~al.(2017a)Kiss, Miller, and Simon]{Kiss_Miller_Simon_2017}
I.~Z. Kiss, J.~C. Miller, and P.~L. Simon.
\newblock Mathematics of epidemics on networks: from exact to approximate models.
\newblock \emph{Springer International Publishing}, 2017a.

\bibitem[Kleinbaum et~al.(1981)Kleinbaum, Morgenstern, and Kupper]{kleinbaum1981selection}
D.~G. Kleinbaum, H.~Morgenstern, and L.~L. Kupper.
\newblock Selection bias in epidemiologic studies.
\newblock \emph{American Journal of Epidemiology}, 1981.

\bibitem[Leskovec and Faloutsos(2006{\natexlab{a}})]{Leskovec_Faloutsos_2006}
J.~Leskovec and C.~Faloutsos.
\newblock Sampling from large graphs.
\newblock In \emph{Proceedings of the 12th ACM SIGKDD International Conference on Knowledge Discovery and Data Mining}, Philadelphia PA USA, 2006{\natexlab{a}}. ACM.

\bibitem[Leskovec and Faloutsos(2006{\natexlab{b}})]{leskovec2006sampling}
J.~Leskovec and C.~Faloutsos.
\newblock Sampling from large graphs.
\newblock In \emph{Proceedings of the 12th ACM SIGKDD International Conference on Knowledge Discovery and Data Mining}, 2006{\natexlab{b}}.

\bibitem[Li et~al.(2005)Li, Alderson, Doyle, and Willinger]{li2005towards}
L.~Li, D.~Alderson, J.~C. Doyle, and W.~Willinger.
\newblock Towards a theory of scale-free graphs: Definition, properties, and implications.
\newblock \emph{Internet Mathematics}, 2005.

\bibitem[Li et~al.(2015)Li, Yu, Qin, Mao, and Jin]{li2015random}
R.~H. Li, J.~X. Yu, L.~Qin, R.~Mao, and T.~Jin.
\newblock On random walk based graph sampling.
\newblock In \emph{2015 IEEE 31st International Conference on Data Engineering}. IEEE, 2015.

\bibitem[Lieberthal and Gardner(2021)]{lieberthal2021connectivity}
Brandon Lieberthal and Allison~M Gardner.
\newblock Connectivity, reproduction number, and mobility interact to determine communities’ epidemiological superspreader potential in a metapopulation network.
\newblock \emph{PLOS Computational Biology}, 2021.

\bibitem[Maiya and Berger-Wolf(2011)]{maiya2011benefits}
A.~S. Maiya and T.~Y. Berger-Wolf.
\newblock Benefits of bias: towards better characterization of network sampling.
\newblock In \emph{Proceedings of the 17th ACM SIGKDD International Conference on Knowledge Discovery and Data Mining}, 2011.

\bibitem[Malmros et~al.(2016)Malmros, Masuda, and Britton]{malmros2016random}
J.~Malmros, N.~Masuda, and T.~Britton.
\newblock Random walks on directed networks: inference and respondent-driven sampling.
\newblock \emph{Journal of Official Statistics}, 2016.

\bibitem[Mann and Whitney(1947)]{mann1947test}
H.~B. Mann and D.~R. Whitney.
\newblock On a test of whether one of two random variables is stochastically larger than the other.
\newblock \emph{The Annals of Mathematical Statistics}, 1947.

\bibitem[Miller and Ting(2020)]{miller2020eon}
J.~C. Miller and T.~Ting.
\newblock Eon (epidemics on networks): a fast, flexible python package for simulation, analytic approximation, and analysis of epidemics on networks.
\newblock \emph{arXiv preprint arXiv:2001.02436}, 2020.

\bibitem[Milligan et~al.(2004)Milligan, Njie, and Bennett]{milligan2004comparison}
P.~Milligan, A.~Njie, and S.~Bennett.
\newblock Comparison of two cluster sampling methods for health surveys in developing countries.
\newblock \emph{International Journal of Epidemiology}, 2004.

\bibitem[Nelson and Williams(2014)]{nelson2014infectious}
K.~E. Nelson and C.~M. Williams.
\newblock \emph{Infectious Disease Epidemiology: Theory and Practice}.
\newblock Jones \& Bartlett Publishers, 2014.

\bibitem[Newman(2018)]{newman2018networks}
M.~Newman.
\newblock \emph{Networks}.
\newblock Oxford University Press, 2018.

\bibitem[Newman and Watts(1999)]{Newman_Watts_1999}
M.~E.~J. Newman and D.~J. Watts.
\newblock Renormalization group analysis of the small-world network model.
\newblock \emph{Physics Letters A}, 1999.

\bibitem[Newman et~al.(2001)Newman, Strogatz, and Watts]{newman2001random}
M.~E.~J. Newman, S.~H. Strogatz, and D.~J. Watts.
\newblock Random graphs with arbitrary degree distributions and their applications.
\newblock \emph{Physical Review E}, 2001.

\bibitem[Nielsen et~al.(2020)Nielsen, Sneppen, Simonsen, and Mathiesen]{nielsen2020heterogeneity}
B.~F. Nielsen, K.~Sneppen, L.~Simonsen, and J.~Mathiesen.
\newblock Heterogeneity is essential for contact tracing.
\newblock \emph{medRxiv}, 2020.

\bibitem[Nunes et~al.(2024)Nunes, Thommes, Fröhlich, Flahault, Arino, Baguelin, Biggerstaff, Bizel-Bizellot, Borchering, Cacciapaglia, et~al.]{nunes2024redefining}
M.~C. Nunes, E.~Thommes, H.~Fröhlich, A.~Flahault, J.~Arino, M.~Baguelin, M.~Biggerstaff, G.~Bizel-Bizellot, R.~Borchering, G.~Cacciapaglia, et~al.
\newblock Redefining pandemic preparedness: multidisciplinary insights from the cerp modelling workshop in infectious diseases, workshop report.
\newblock \emph{Infectious Disease Modelling}, 2024.

\bibitem[Paton et~al.(2018)Paton, Gubbins, and King]{paton2018understanding}
David~J Paton, Simon Gubbins, and Donald~P King.
\newblock Understanding the transmission of foot-and-mouth disease virus at different scales.
\newblock \emph{Current opinion in virology}, 2018.

\bibitem[Qi et~al.(2023)Qi, Tan, Chen, Duan, and Lu]{qi2023efficient}
M.~Qi, S.~Tan, P.~Chen, X.~Duan, and X.~Lu.
\newblock Efficient network intervention with sampling information.
\newblock \emph{Chaos, Solitons \& Fractals}, 2023.

\bibitem[Richardson et~al.(2003)Richardson, Agrawal, and Domingos]{richardson2003trust}
M.~Richardson, R.~Agrawal, and P.~Domingos.
\newblock Trust management for the semantic web.
\newblock In \emph{International semantic Web conference}. Springer, 2003.

\bibitem[Rosvall and Bergstrom(2008)]{rosvall2008maps}
M.~Rosvall and C.~T. Bergstrom.
\newblock Maps of random walks on complex networks reveal community structure.
\newblock \emph{Proceedings of the National Academy of Sciences}, 2008.

\bibitem[SeyedAlinaghi et~al.(2021)SeyedAlinaghi, Abbasian, Solduzian, Yazdi, Jafari, Adibimehr, Farahani, Khaneshan, Alavijeh, Jahani, et~al.]{seyedalinaghi2021predictors}
S.~A. SeyedAlinaghi, L.~Abbasian, M.~Solduzian, N.~Ayoobi Yazdi, F.~Jafari, A.~Adibimehr, A.~Farahani, A.~S. Khaneshan, P.~Ebrahimi Alavijeh, Z.~Jahani, et~al.
\newblock Predictors of the prolonged recovery period in covid-19 patients: a cross-sectional study.
\newblock \emph{European Journal of Medical Research}, 2021.

\bibitem[Shen et~al.(2009)Shen, Ning, and Qin]{shen2009analyzing}
Yu~Shen, Jing Ning, and Jing Qin.
\newblock Analyzing length-biased data with semiparametric transformation and accelerated failure time models.
\newblock \emph{Journal of the American Statistical Association}, 2009.

\bibitem[Spencer(2021)]{spencer2021accelerating}
S.~E.~F. Spencer.
\newblock Accelerating adaptation in the adaptive metropolis--hastings random walk algorithm.
\newblock \emph{Australian \& New Zealand Journal of Statistics}, 2021.

\bibitem[Stein et~al.(2014)Stein, Steenbergen, Chanyasanha, Tipayamongkholgul, Buskens, Heijden, Sabaiwan, Bengtsson, Lu, Thorson, et~al.]{stein2014online}
M.~L. Stein, J.~E.~Van Steenbergen, C.~Chanyasanha, M.~Tipayamongkholgul, V.~Buskens, P.~G. M. Van~Der Heijden, W.~Sabaiwan, L.~Bengtsson, X.~Lu, A.~E. Thorson, et~al.
\newblock Online respondent-driven sampling for studying contact patterns relevant for the spread of close-contact pathogens: a pilot study in thailand.
\newblock \emph{PloS One}, 2014.

\bibitem[Suhail et~al.(2021)Suhail, Afzal, and Kshitiz]{suhail2021incorporating}
Yasir Suhail, Junaid Afzal, and Kshitiz.
\newblock Incorporating and addressing testing bias within estimates of epidemic dynamics for sars-cov-2.
\newblock \emph{BMC medical research methodology}, 2021.

\bibitem[Tyrer and Heyman(2016)]{tyrer2016sampling}
S.~Tyrer and B.~Heyman.
\newblock Sampling in epidemiological research: issues, hazards and pitfalls.
\newblock \emph{BJPsych Bulletin}, 2016.

\bibitem[{University of Oregon}(2025)]{oregon_route}
{University of Oregon}.
\newblock Route views project, 2025.
\newblock URL \url{https://www.routeviews.org/bkup.index.html}.
\newblock Accessed: 2025.

\bibitem[Vynnycky and White(2010)]{vynnycky2010introduction}
E.~Vynnycky and R.~White.
\newblock An introduction to infectious disease modelling.
\newblock \emph{OUP Oxford}, 2010.

\bibitem[Watts and Strogatz(1998)]{watts1998collective}
D.~J. Watts and S.~H. Strogatz.
\newblock Collective dynamics of ‘small-world’ networks.
\newblock \emph{Nature}, 1998.

\bibitem[Wei et~al.(2004)Wei, Erenrich, and Selman]{wei2004towards}
W.~Wei, J.~Erenrich, and B.~Selman.
\newblock Towards efficient sampling: exploiting random walk strategies.
\newblock In \emph{AAAI}, 2004.

\bibitem[{World Organisation for Animal Health}(2024)]{woah_fmd}
{World Organisation for Animal Health}.
\newblock Foot-and-mouth disease, 2024.
\newblock URL \url{https://www.woah.org/en/disease/foot-and-mouth-disease/}.
\newblock Accessed: 2024.

\bibitem[Zhang et~al.(2021)Zhang, Lin, and Zhang]{zhang2021discovering}
S.~Zhang, X.~Lin, and X.~Zhang.
\newblock Discovering dti and ddi by knowledge graph with mhrw and improved neural network.
\newblock In \emph{2021 IEEE International Conference on Bioinformatics and Biomedicine (BIBM)}. IEEE, 2021.

\end{thebibliography}





\end{document}